\documentclass[acmsmall]{acmart}  

\AtBeginDocument{%
  \providecommand\BibTeX{{%
    \normalfont B\kern-0.5em{\scshape i\kern-0.25em b}\kern-0.8em\TeX}}}

\usepackage{multirow}
\usepackage{bbding}
\usepackage{subfig}
\usepackage{bm}

\usepackage{dsfont}
\usepackage{threeparttable}
\usepackage{amsmath}
\usepackage{amsfonts}
\usepackage{color}
\usepackage{comment}
\usepackage{graphicx}
\usepackage{multirow}

\usepackage{url}
\usepackage{booktabs}  
\usepackage{enumitem}
\usepackage{float}
\usepackage{array}
\usepackage{float}

\usepackage{algorithm}
\usepackage{algorithmic}
\usepackage[switch]{lineno}

\usepackage{lscape}
\usepackage{booktabs}
\usepackage{multicol}

\begin{document}

\title{Understanding Before Recommendation: Semantic Aspect-Aware Review Exploitation via Large Language Models}

\author{Fan Liu}
\orcid{0000-0002-4547-3982}
\affiliation{%
  \institution{National University of Singapore}
  \streetaddress{School of Computing, 13 Computing Drive}
  \city{Singapore}
  \country{Singapore}
  \postcode{117417}
}
\email{liufancs@gmail.com}

\author{Yaqi Liu}
\affiliation{
  \institution{National University of Singapore}
  \streetaddress{School of Computing, 13 Computing Drive}
  \city{Singapore}
  \country{Singapore}
  \postcode{117417}
}
\email{liuyaqics@gmail.com}

\author{Huilin Chen}
\email{ClownClumsy@outlook.com}
\affiliation{%
  \institution{Hefei University of Technology}
  \streetaddress{193 Tun Xi Lu, Baohe District}
  \city{Hefei}
  \postcode{230002}
  \country{China}
  }

\author{Zhiyong Cheng}
\email{jason.zy.cheng@gmail.com}
\affiliation{%
  \institution{Hefei University of Technology}
  \streetaddress{193 Tun Xi Lu, Baohe District}
  \city{Hefei}
  \postcode{230002}
  \country{China}
  }

\author{Liqiang Nie}
\affiliation{%
  \institution{Harbin Institute of Technology, Shenzhen}
  \streetaddress{School of Computer Science and Technology, HIT Campus of University Town of Shenzhen}
  \city{Shenzhen}
  \country{China}
  \postcode{518055}
}
\email{nieliqiang@gmail.com}

\author{Mohan Kankanhalli}
\affiliation{%
 \institution{National University of Singapore}
 \streetaddress{School of Computing, 13 Computing Drive}
 \city{Singapore}
 \country{Singapore}
 \postcode{117417}
 }
 \email{mohan@comp.nus.edu.sg}

\begin{abstract}
Recommendation systems harness user-item interactions like clicks and reviews to learn their representations. 
Previous studies improve recommendation accuracy and interpretability by modeling user preferences across various aspects and intents.
However, the aspects and intents are inferred directly from user reviews or behavior patterns, suffering from the data noise and the data sparsity problem. Furthermore, it is difficult to understand the reasons behind recommendations due to the challenges of interpreting implicit aspects and intents.
To address these constraints, we harness the sentiment analysis capabilities of Large Language Models (LLMs) to enhance the accuracy and interpretability of the conventional recommendation methods. 
Specifically, inspired by the deep semantic understanding offered by LLMs, we introduce a chain-based prompting strategy to uncover semantic aspect-aware interactions, which provide clearer insights into user behaviors at a fine-grained semantic level. To incorporate the rich interactions of various aspects, we propose the simple yet effective Semantic Aspect-based Graph Convolution Network (short for SAGCN). By performing graph convolutions on multiple semantic aspect graphs, SAGCN efficiently combines embeddings across multiple semantic aspects for final user and item representations. The effectiveness of the SAGCN was evaluated on four publicly available datasets through extensive experiments, which revealed that it outperforms all other competitors. Furthermore, interpretability analysis experiments were conducted to demonstrate the interpretability of incorporating semantic aspects into the model.
\end{abstract}

\ccsdesc[500]{Information systems~Personalization}
\ccsdesc[500]{Information systems~Recommender systems}
\ccsdesc[500]{Information systems~Collaborative filtering}

\keywords{Recommendation, Semantic Aspect, Large Language Models, Collaborative Filtering}

\maketitle
\section{Introduction}
Recommendation systems have become essential components of various online platforms. They not only assist users in navigating through overwhelming volumes of information but also serve as a powerful tool for service providers to increase customer retention and revenue. Within this field, Collaborative Filtering (CF)-based models have achieved significant advancements in learning the user and item representations by effectively exploiting the historical user-item interactions~\cite{netflix,he2017neural,wu2016cdae,xue2017deep}. For instance, Matrix Factorization (MF) decomposes the user-item interaction matrices into lower-dimensional latent vectors and predicts the user preference with the inner product~\cite{netflix}. By replacing the inner product with nonlinear neural networks, NeuMF~\cite{he2017neural} can model the complex interaction behaviors, resulting in better recommendation performance. 
More recently, Graph Convolution Networks (GCNs)-based models~\cite{berg2019gcmc,wang2019ngcf,wang2019kdd,He@LightGCN,wang2020DGCF,Liu2021IMP_GCN,fan2022GTN} offer a cutting-edge approach in recommendation systems by leveraging the high-order connectivities based on the topological structure of the interaction data. For example, NGCF~\cite{wang2019ngcf} incorporates high-order connectivities between user and item nodes in the bipartite graph to learn their representations for recommendation.

To better understand user preferences, researchers have paid significant attention to exploiting fine-grained behaviors between users and items on different aspects or intents~\cite{cheng2018aspect,cheng20183ncf,chin2018anr,liu2018MAML,Chen2024CDR}. Existing methods can be categorized into two main approaches. First, employing topic models~\cite{blei2003latent}, the statistical model designed to identify the abstract ``topics'' based on the frequency and co-occurrence of words in textual content, to extract aspects from these reviews. These extracted aspects are then incorporated into user preference modeling~\cite{cheng2018aspect, cheng20183ncf}.
In particular, ALFM~\cite{cheng2018aspect} applies a topic model on reviews to extract the topics the user pays attention to, while A$^3$NCF~\cite{cheng20183ncf} learns the user attention from reviews via a neural attention network in an end-to-end manner. While these models can identify topics or aspects in the review, understanding their semantic meaning is not straightforward. In addition, the noise in the review has a negative impact on the aspect extraction. For example, words that have no contribution to user preference, like ``the'', ``last'', and ``year'' in reviews, are extracted by the topic models and leveraged into the training process of the recommendation model.
Second, disentangling the underlying aspect/factor behind interaction data~\cite{wang2020DisenHAN,wang2020DGCF, Liu2022DMRL, ADDRL@MM}. Disentangled representations are becoming increasingly important for learning robust and independent representations from user-item interaction data in recommendation~\cite{wang2020DisenHAN,wang2020DGCF}. For example, DisenHAN~\cite{wang2020DisenHAN} learns disentangled user/item representations from different aspects in a heterogeneous information network, while DGCF~\cite{wang2020DGCF} models a distribution over intents for each user-item interaction to study the diversity of user intents on adopting the items. Despite their ability to accurately model user preference and capture the underlying factors, 
such models extract the aspects relying on user-item interaction data, which suffer from data sparsity and bias issues in recommendations. In addition, the implicit aspects and intents behind the recommendations are often elusive and not readily interpretable.

\begin{figure}[t]
		\centering 
		\includegraphics[width=0.55\linewidth]{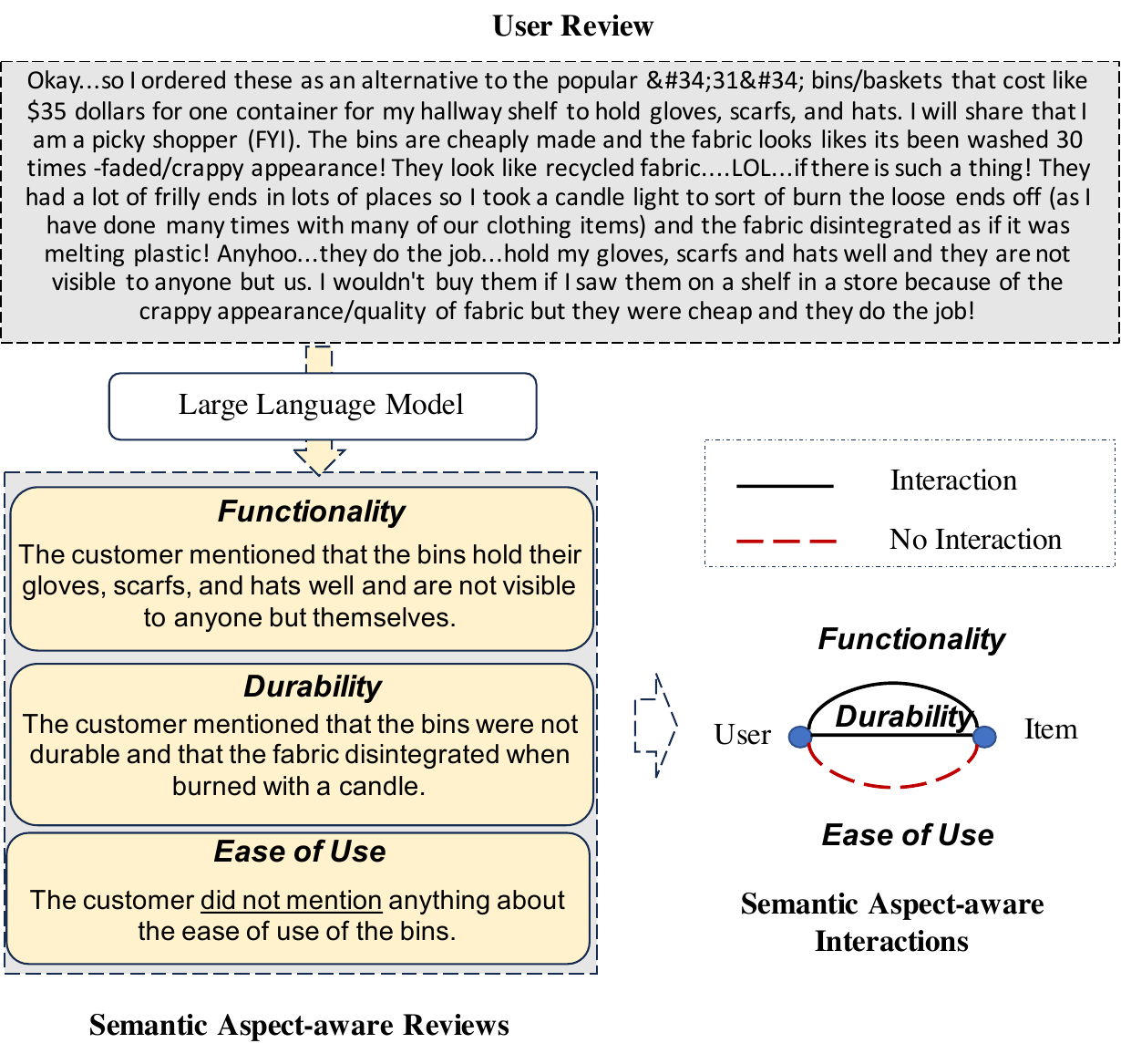} 
		\caption{An example of extracting semantic aspect-aware reviews (e.g., functionality, durability, and ease of use) from user review via LLM. Based on the extracted reviews, the semantic aspect-aware interactions are discerned. The results from LLM indicate that user $u$ interacts with item $i$ in terms of functionality and durability aspects, but dose not interacts it in relation to ease of use.}
		\label{fig:alphacompare}  
		\vspace{0.0cm}
\end{figure}

Semantic aspects often denote specific features or attributes of a product or service in which users might be interested in the context of recommendation systems and sentiment analysis. 
Extracting semantic aspects from reviews could improve our comprehension of the complex relations between each user and item at a finer level. Moreover,
semantic aspects, like meaningful attributes or concepts of an item, provide a deeper understanding of its inherent features, thereby enhancing the interpretability of recommendations. As shown in Fig.~\ref{fig:alphacompare}, the review might mention various semantic aspects such as \emph{functionality}, \emph{durability}, and \emph{ease of use}. Each of these aspects carries a specific meaning related to user experience. 
By analyzing and summarizing the raw review into structured semantic aspects, we can better understand the underlying rationales behind user behaviors. Focusing on reviews that highlight specific item aspects enables us to discern semantic aspect-aware interactions. For instance, a review might emphasize aspects like \emph{functionality} and \emph{durability} but omit \emph{ease of use}. 
Such patterns indicate distinct interactions between the user and item concerning \emph{functionality} and \emph{durability}, but not for \emph{ease of use}. 
Utilizing these semantic-aware insights leads to richer and more accurate representations of both users and items.

Large Language Models (LLMs) have emerged as powerful tools in natural language processing and have recently gained significant attention in various fields~\cite{Da2020AIR, Sileo2022LLM}. Their contextual understanding allows for distilling various semantic aspects based on reviews without being affected by noise present in reviews. In this paper, we propose a novel approach that combines the strengths of LLMs with conventional recommendation techniques by extracting the semantic aspects from user reviews.
Although LLMs exhibit advanced proficiency across numerous tasks, they still encounter challenges in accurately identifying aspects, especially in autonomously discovering unseen aspects without explicit instructions. 
{To mitigate this constraint of LLMs, we devise a chain-based prompting strategy, which consists of the \emph{semantic aspects extraction prompt} and \emph{semantic aspects-aware review extraction prompt}. The first prompt extracts semantic aspects from raw reviews, after which we refine and select high-quality aspects from the entire dataset. The second prompt then identifies reviews that align with these high-quality semantic aspects.} On the basis of the extracted aspect-aware reviews, we can form aspect-specific interactions, capturing users' preferences from various semantic aspects. To demonstrate the potential of structuring raw reviews into semantic aspects via LLMs in recommendations, we present a \textbf{S}emantic \textbf{A}spect-based \textbf{G}raph \textbf{C}onvolution \textbf{N}etwork (SAGCN) model. This model performs graph convolution
operations on aspect-aware interaction graphs, each of which is constructed based on the user-item interactions of a specific aspect derived from the aspect-aware reviews. The final user and item representations are obtained by combining the embeddings learned from different aspect-specific graphs. The merits of our model are twofold: (1) embeddings are enriched by learning from various aspects; (2) interpretability is improved by incorporating semantic aspects. Extensive experiments have been conducted on three benchmark datasets to evaluate the SAGCN. The results indicate that SAGCN outperforms state-of-the-art GCN-based recommendation models in generating more effective user and item representations. Additionally, interpretability analysis experiments demonstrate the effectiveness of integrating semantic aspects into our model. To facilitate further research, we have released the code and the constructed semantic aspect-aware interaction data for others to replicate this work~\footnote{https://github.com/HuilinChenJN/LLMSAGCN.}.


In summary, the main contributions of this work are as follows:
\begin{itemize}[leftmargin=*]

\item We emphasize the significance of utilizing semantic aspect-aware interaction in recommendations. Inspired by this, we propose a novel approach that combines the strengths of LLMs with conventional recommendation techniques by extracting the semantic aspects from user reviews.

\item To extract the semantic aspects and semantic aspect-aware reviews, we devise a chain-based prompting strategy. On the basis of this, we could discern aspect-aware interactions within each user-item review.

\item To demonstrate the effectiveness of our approach in enhancing recommendation accuracy and interoperability, we present the SAGCN model that leverages the extracted semantic aspect-aware interactions for user preference modeling.

\item {We conducted empirical studies on four benchmark datasets to evaluate the performance of SAGCN. Our results demonstrate that SAGCN outperforms state-of-the-art recommendation models in both accuracy and interpretability.}

\end{itemize}

\section{related work}
\label{sec:relatedwork}
This section briefly introduces recent advances in collaborative filtering based on user reviews, GCN-based recommendation models, and large language models in recommendation, which are closely relevant to our work. 
\subsection{Collaborative Filtering Based on User Reviews}
Collaborative Filtering (CF) techniques~\cite{Koren2009MF,rendle2009bpr,wang2019kdd,he2017neural,wang2019ngcf,He@LightGCN,liu2018MAML} play a pivotal role in recommendation systems. 
The fundamental concept involves learning representations (or embedding vectors) of users and items by harnessing collaborative information derived from user-item interactions. These learned user and item embedding vectors are then utilized within an interaction function to predict a user's preference for an item.
For example, Matrix Factorization (MF)~\cite{Koren2009MF} learns the user and item embeddings by minimizing the error of re-constructing the user-item interaction matrix. Due to the powerful capabilities of deep learning, it is widely adopted in recommendations to improve the quality of user and item embeddings~\cite{xue2017deep} or model more complicated interactions between users and items~\cite{he2017neural, Li2019CNR}. One prominent method is  NeuMF~\cite{he2017neural}, which leverages a nonlinear neural network to model complex interactions between users and items. 

To mitigate the data sparsity problem, user reviews~\cite{catherine2017transnets,mcauley2013hidden,tan2016rating,cheng20183ncf,chen2018neural,cheng2018aspect,chin2018anr,Wei2023LightGT,Han@MetaMMRS} have been leveraged in recommender systems to learn the user and item representations as they can provide additional information about user preference. One of the typical methods to learn user preferences by exploiting reviews is through deep learning techniques~\cite{DeepCoNN2017WSDM}. For example, DeepCoNN~\cite{DeepCoNN2017WSDM} learn item characteristics and user preferences jointly from user reviews via the deep model. Existing methods extract aspects from reviews and integrate these aspects into user preference modeling. For example, A$^3$NCF~\cite{cheng20183ncf} extracts user preferences and item characteristics from review texts via topic model to enhance the user's and item's representations. {SULM~\cite{Bauman2017SIGKDD} uses a sentiment analysis system, Opinion Parser, to extract aspects. Through a matrix factorization approach, it calculates the personal impact each aspect has on the user's overall rating.}
Furthermore, it is well-recognized that multimodal features can provide complementary information to each other, as demonstrated in~\cite{zhang2016collaborative,zhang2017joint,Bauman2017SIGKDD,liu2018MAML}.
JRL~\cite{zhang2017joint} first utilizes deep neural networks to separately extract user and item features from multiple modalities, such as ratings, reviews, and images, then forms the final user and item representations by concatenating those multimodal features. To model users' diverse preferences, MAML~\cite{liu2018MAML} fuses user reviews with item images to estimate the weights for different aspects of items. 

Despite their success, the above-mentioned methods are still incapable of computing the optimal embeddings for recommendation. This deficiency arises from the fact that collaborative filtering signals are only implicitly captured, thereby neglecting the transitivity property of behavioral similarity~\cite{wang2019ngcf}. A common solution is to explicitly incorporate the graph structure to enhance user and item representation learning. In the next, we mainly discussed the Graph-based recommendation methods.

\subsection{Large Language Models in Recommendation}
Large Language Models (LLMs) have emerged as potent tools in natural language processing, garnering increasing attention in recommendation systems. 
The LLM-based models excel in capturing contextual information and comprehending user queries, item descriptions, and other textual data in a highly effective manner~\cite{RLP2022RecSys, Da2020AIR, Sileo2022LLM}. By understanding the context, LLM-based models can improve the recommendation performance. 

There are two primary approaches to integrating LLMs into recommendation systems.
Firstly, LLMs can be adapted directly as recommendation models. Many works in this vein treat recommendation tasks as natural language tasks, employing techniques such as in-context learning, prompt tuning, and instruction tuning to tailor LLMs for generating recommendation results.~\cite{Ji2023GenRec, Zhang2023LLM,Zhang2023chatgpt,wang2023LLM,Li2023PEPLER}. Recent attention has been particularly drawn to this line of research, buoyed by the impressive capabilities showcased by models like ChatGPT. For instance, Ji et al.~\cite{Ji2023GenRec} proposed a generative recommendation method that uses LLMs to generate recommended items. {PEPLER~\cite{Li2023PEPLER} generates natural language explanations for recommendations by treating user and item IDs as prompts.}
Secondly, LLMs can be leveraged to extract high-quality representations of textual features or tap into the extensive external knowledge they encapsulate~\cite{Liu2023Newsgeneration}. 
Most existing efforts align pre-trained models like BERT~\cite{Devlin2019BERTPO} with domain-specific data through fine-tuning. Additionally, beyond serving as standalone recommendation systems, some studies employ LLMs to construct model features. 
For example, GENRE~\cite{Liu2023Newsgeneration} introduced three prompts to conduct three feature enhancement sub-tasks for news recommendation. The prompts involved refining news titles using ChatGPT, extracting profile keywords from user reading history, and generating synthetic news to enhance user interactions based on their history. By incorporating these LLM-constructed features, traditional news recommendation models can be greatly improved. {RLMRec~\cite{ren2024representation} generates user and item profiles based on interaction data and textual information, such as product descriptions, user reviews, and product titles. Then, these generated profiles are encoded as semantic information of LLMs and incorporated into the CF-based recommendation system.} 

Utilizing LLM's sentiment analysis capability, we aim to better understand users' various intentions expressed in reviews. Specifically, we first extract semantic aspects from user reviews and then identify the semantic-aspect interactions for each user and item pair. By leveraging semantic-aspect interactions, we can capture fine-grained user preferences. Compared to conventional methods that use features directly extracted from reviews by pretrained language models, semantic-aware interactions can improve the interpretability of recommendation models.

\subsection{GCN-based Recommendation Models}
Early recommendation approaches employ graph-based approaches to infer indirect user-item preferences. The high-order proximity between users and items is leveraged through random walks on the graph in order to provide effective recommendations~\cite{gori2007itemrank,fouss2007tkde}. The recently proposed approaches exploit the user-item interaction graph to enrich the user-item interactions~\cite{hoprec,yu2018walkranker} as well as to explore additional collaborative relations, such as similarities between users and items~\cite{chen2019cse,yu2018walkranker}. For example, WalkRanker~\cite{yu2018walkranker} and CSE~\cite{chen2019cse} utilize random walks to investigate the high-order proximity in both user-user and item-item relationships. These methods rely on random walks to sample interactions for model training, thus their performance is heavily depends on the quality of the generated interactions. 

In recent years, Graph Convolutional Networks (GCNs) have garnered significant attention in the realm of recommendation systems, owing to their robust capabilities in learning node representations from non-Euclidean structures~\cite{berg2019gcmc,wang2019ngcf,ying2018pinsage,wang2019kdd,wang2020DGCF, He@LightGCN, Liu2021IMP_GCN, lin2022NCL, Cheng2023MBR, Ren2023DCCF,ClusterGCF2024TOIS}. This also led to the rapid advancement of GCN-based recommendation models. For instance, NGCF~\cite{wang2019ngcf} propagates embeddings on the historical interaction graph to exploit the high-order proximity. Some research has been dedicated to simplifying GCN (graph convolutional network) and GCN-based recommendation models~\cite{pmlr-v97-wu19e}. As a typical method, He et al.~\cite{He@LightGCN} observed that the two common designs feature transformation and nonlinear activation have no positive effects on the final performance. Hence, they proposed LightGCN, which removes these two modules and significantly improves recommendation performance.
To obtain robust user and item representations, fine-grained interactions are exploited to refine the graph structure. Wang et al.~\cite{wang2020DGCF} presented a GCN-based model, which yields disentangled representations by modeling a distribution over intents for each user-item interaction; 
IMP-GCN~\cite{Liu2021IMP_GCN} learns user and item embeddings by performing high-order graph convolution inside subgraphs constructed by the user nodes with common interests and associated items. By incorporating side information, GRCN~\cite{Wei2019GRCN} modify the structure of the interaction graph adaptively by discovering and pruning potential false-positive edges; and
LATTICE~\cite{LATTICE2021TMM} mines the latent structure between items by learning an item-item graph. Graph convolutional operations are performed on both item-item and user-item interaction graphs to learn user and item representations.

In this paper, we present a simple yet effective GCN-based recommendation model that propagates embeddings over semantic aspect-based user-item interaction graphs. These graphs are built using the semantic aspect-aware interactions extracted by LLM from user reviews. Our proposed model outperforms existing GCN-based recommendation methods in terms of accuracy and interoperability.

\section{Methodology}
\label{Sec:method}
\subsection{Preliminary}
\subsubsection{Problem Setting and Notations Definition.}
To provide the details for our approach, let's first outline the problem setting. We are working with a dataset that consists of an interaction matrix denoted as $\bm{R}^{N \times M}$, where $\mathcal{U}$ represents the set of users, and $\mathcal{I}$ represents the set of items. $N$ and $M$ represent the total number of users and items, respectively. The matrix $\bm{R}$ records the interaction history between users and items. A nonzero entry $r_{ui}\in \bm{R}$ indicates that user $u \in \mathcal{U}$  has interacted with item ${i} \in \mathcal{I}$ before; otherwise, the entry is zero. For each user-item pair $(u, i)$, a review $v$ is associated with them.
{In this work, our goal is to explore the potential of reviews to enhance recommendation accuracy and interpretability.}

To achieve this goal, we propose an approach that integrates LLMs with conventional recommendation techniques. Specifically, we (1) design a chain-based prompting approach, which can effectively extract semantic aspect-aware reviews without clear instructions; (2) introduce a semantic aspect-based GCN model for recommendation. We use $\{\bm{R^1}, \bm{R^2}, \cdots, \bm{R^{A}}\}$ to denote the semantic aspect-based interaction matrices. $A$ is the number of aspects. $\bm{R^a}$ is the interaction matrix of the $a$-th aspect.

\subsection{Overview}
Recommendation systems learn user and item representations by exploiting user-item interactions, such as clicks and reviews. In this way, users' implicit preferences are encoded into latent vectors. Notably, user reviews offer explicit intents of a user towards an item. For example, the review might mention various semantic aspects such as \emph{functionality}, \emph{durability}, and \emph{ease of use}. Each of these aspects carries a specific meaning related to user experience. Recognizing these explicit intentions leads to a deeper understanding of user preferences. More recently, LLMs have emerged as powerful tools for sentiment analysis~\cite{Sileo2022LLM}, capable of extracting semantic aspects from user reviews. 
Therefore, to enhance the accuracy of recommendation systems, a desired approach should well seamlessly combine the advanced capabilities of LLMs with conventional recommendation techniques.

\begin{figure*}[t]
		\centering 
		\includegraphics[width=0.6\linewidth]{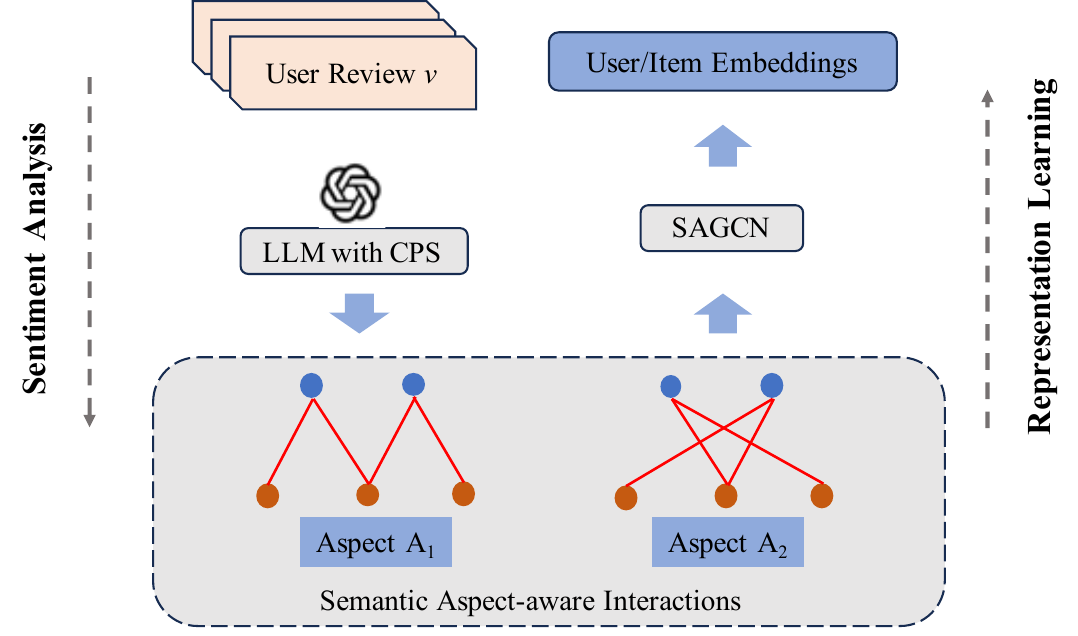} 
		\caption{Overview of Our Approach. Our approach consists of two components: sentiment analysis and representation learning. The former leverages the LLM with a Chain-based Prompting Strategy (short for CPS) to extract the semantic aspect-aware interactions from user reviews. The latter is dedicated to learning user and item representations based on these interactions, employing the Semantic Aspect-aware Graph Convolutional Network (short for SAGCN).}
		\label{fig:approach}
		\vspace{0.0cm}
\end{figure*}

In this paper, we explore the fusion of the LLMs with conventional recommendation techniques to enhance both the accuracy and interoperability of recommendations. 
As shown in Fig.~\ref{fig:approach}, our approach begins with the sentiment analysis of user-item reviews, utilizing LLMs with the chain-based prompting strategy to extract semantic aspect-aware interactions from each review.
These extracted semantic aspect-aware reviews shed light on the fine-grained interactions underlying each user-item relationship. On the basis of these insights, we introduce a semantic aspect-based graph convolution network to enhance the user and item representations by leveraging these semantic aspect-aware interactions.
To this end, our approach can properly extract and leverage the semantic aspect-aware interactions among each user-item pair for better user and item representation learning.

Next, we detail our method for semantic aspect-aware review extraction and semantic aspect-based graph convolution network.

\subsection{Chain-based Prompting Strategy}

In this section, we conduct sentiment analysis on user reviews using LLM to discern user-item interactions on various semantic aspects. Despite the powerful capability of LLMs in sentiment analysis, it is still a challenge to accurately identify the semantic aspect-aware interactions in each review without any instruction. This is because user reviews often contain complex, nuanced language that varies widely in terms of vocabulary, syntax, and semantics. This complexity makes it difficult to consistently identify specific aspects and associated sentiments. On the other side, while LLMs are adept at grasping context, the depth and breadth of their understanding can be limited by the training data. They might not always capture the full context or the subtle cues that indicate sentiment towards specific aspects, especially when these aspects are not explicitly mentioned. To alleviate these problems, as depicted in Fig.~\ref{fig:prompting}, we introduce a chain-based prompting strategy that begins by collecting the semantic aspects from all user reviews. {Subsequently, we refine and select high-quality semantic aspects from all reviews and use these as the instruction information to generate the semantic aspect-aware reviews for each review.} Finally, we identify aspect-aware interactions based on the generated reviews. 

\subsubsection{Semantic Aspects Extraction}
\begin{figure*}[t]
		\centering 
		\includegraphics[width=1.0\linewidth]{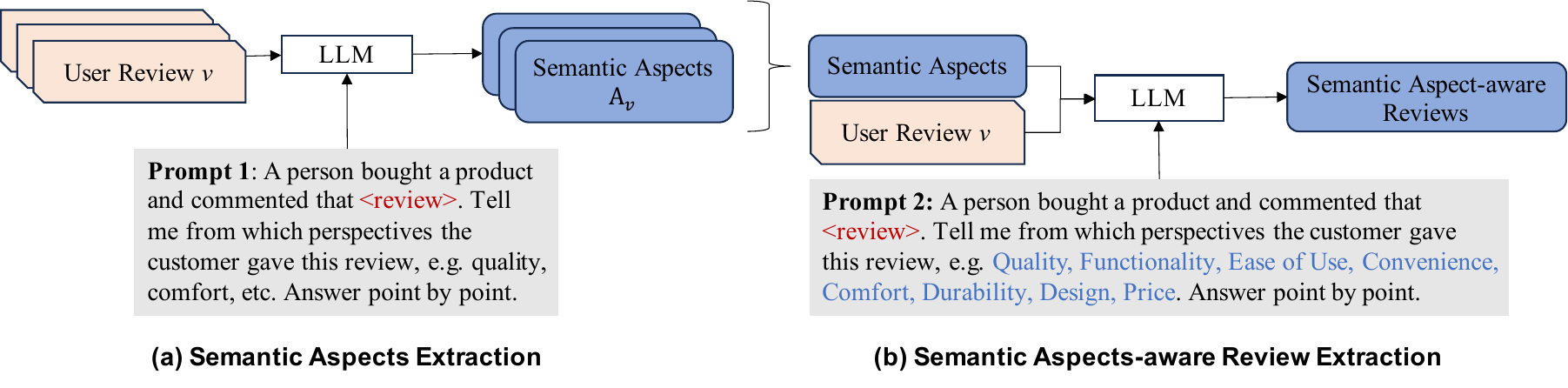} 
		\caption{Semantic Aspects and Semantic Aspect-aware Reviews Extraction.}
		\label{fig:prompting}  
		\vspace{0.0cm}
\end{figure*}
{The semantic aspects vary depending on the recommendation scenario. For instance, when buying dresses, \emph{style} is crucial, but for office products, \emph{ease of use} is the main concern. In this work, we first utilize the understanding capabilities of LLMs to extract unseen semantic aspects from the dataset. }
As illustrated in Fig.~\ref{fig:prompting}(a), the semantic aspects extraction prompt is designed as follows:
\begin{itemize}[leftmargin=*]
\item \emph{Prompt 1}: \emph{A person bought a product and commented that \textcolor{red}{<review>}. Tell me from which perspectives the customer gave this review, e.g., quality, comfort, etc. Answer point by point.} 
\end{itemize}
{Where $<\text{review}>$ denotes a user review, let the review $v$ be analyzed by a LLM to generate a set of semantic aspects $\mathcal{A}_v = \{a_v^1, a_v^2, \ldots, a_v^{N_v} \}$, where $N_v$ is the number of extracted aspects.} It is important to note that the prompt provides minimal information about these semantic aspects. We utilize \emph{quality} and \emph{comfort} as indicative cues for the LLM to identify the unseen semantic aspects within the reviews.
Through the analysis of each review by the LLM, we can compile a comprehensive list of semantic aspects pertinent to the domain of recommendation by amalgamating the semantic aspect sets derived from all reviews. Upon assembling a complete set of semantic aspects, we conduct a statistical examination to prioritize these aspects. This process involves ranking them from highest to lowest frequency and manually consolidating or eliminating aspects with overlapping meanings. The final set of semantic aspects for this recommendation scenario is denoted as $\mathcal{A} = \{a^1, a^2, \ldots, a^N \}$, $N$ is the number of obtained semantic aspects.

\subsubsection{Semantic Aspect-Aware Review Extraction}
{Empirically, LLMs tend to generate more accurate results when provided with detailed prompts. To accurately identify the semantic aspects within each review, we further employ LLM to generate high-quality semantic aspect-aware reviews based on the obtained semantic aspect set $\mathcal{A}$.} As shown in Fig.~\ref{fig:alphacompare}, an aspect-aware review provides a summary that identifies the presence of specific aspects within a review. Since we can focus on specific aspects, high-quality semantic aspect-aware reviews can be obtained. For instance, on the Baby Dataset, we get a semantic aspect set containing \emph{Quality}, \emph{Comfort}, \emph{Durability}, \emph{Design}, \emph{Functionality}, \emph{Ease of use}, \emph{Price}, and \emph{Size}. As illustrated in Fig.~\ref{fig:prompting}(b), we adapt the previous prompt to specific semantic aspects:
\begin{itemize}[leftmargin=*]
\item \emph{Prompt 2}: \emph{A person bought a product and commented that \textcolor{red}{<review>}. Tell me from which perspectives the customer gave this review, e.g., \textcolor{blue}{\{$a^1, a^2, \cdots, a^N$\}}. Answer point by point.} 
\end{itemize}

The LLM identifies the semantic aspect-aware review for the aspects we input. Then, we identify the semantic aspect-aware interactions between the user and the item based on the generated semantic-aware reviews. For example, if the LLM generates the output: ``\emph{The customer mentioned that the bins were not durable and that the fabric disintegrated when burned with a candle.}'', it indicates an interaction on \emph{Functionality}. On the other hand, if the LLM generates the output:
``\emph{The customer did not mention anything about the ease of use of the bins.}'', which means that there is no interaction in this aspect. {Note that the time costs for the Chain-based Prompting strategy vary based on the length of input and output tokens. Based on our empirical evaluations, the process takes approximately 4 to 10 seconds per review.} 

\subsection{Semantic Aspect-Based Graph Convolution Network}

\begin{figure}[t]
		\centering 
		\includegraphics[width=0.6\linewidth]{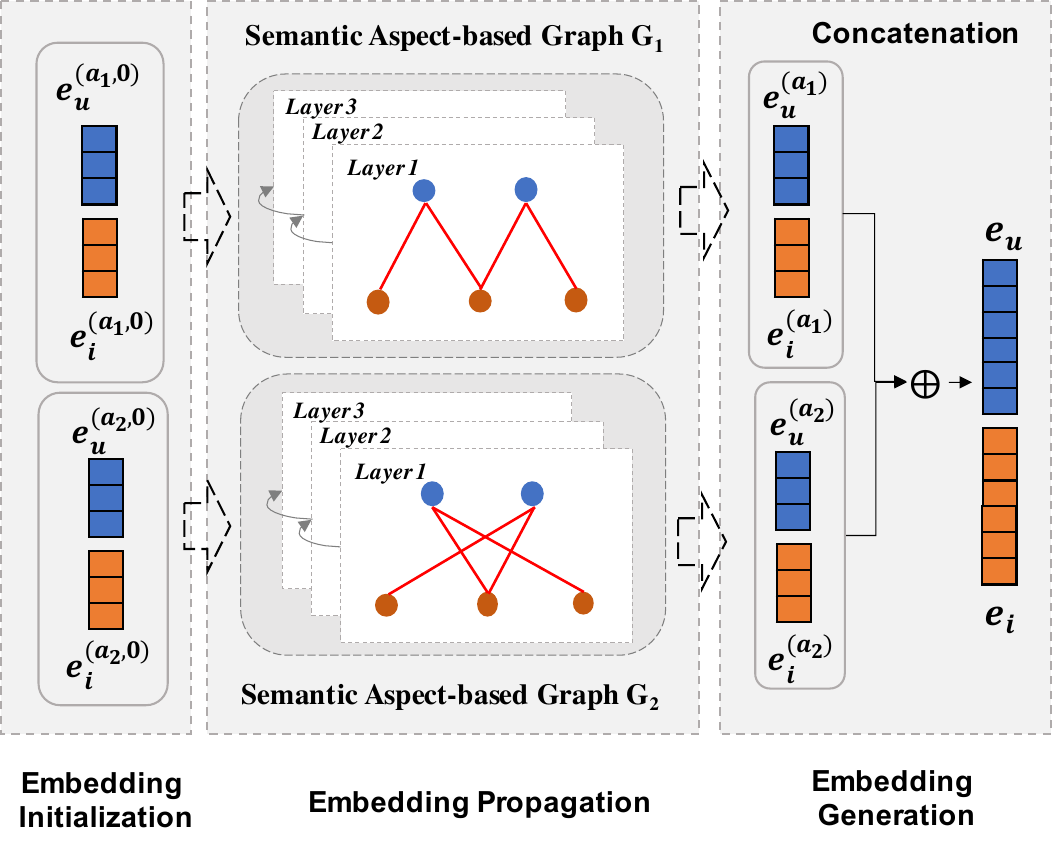} 
		\caption{Overview of our SAGCN model.}
		\label{fig:SAGCN}  
		\vspace{0.0cm}
\end{figure}

Based on the extracted semantic-aware user-item interactions for semantic aspects, we seek to model users' preferences on various aspects with GCN-based techniques. 
The structure of our SAGCN is illustrated in Fig.~\ref{fig:SAGCN}, and it is composed of three key components: 1) \textbf{embedding initialization}, which initialized node (users and items) embeddings for the subsequent learning in each aspect-specific interaction graph; 2) \textbf{embedding propagation}, in which the GCN technique~\cite{He@LightGCN} is adopted to learn the user and item embeddings by performing graph convolution on each semantic aspect-based graph; 3) \textbf{embedding generation}, which aggregates the embeddings learned from each aspect-specific graph and combines the embeddings of different aspects.

\subsubsection{Semantic Aspect-based Graph Construction} 
The semantic aspect-based graph construction is built on the results of aspect-aware review extraction, which is achieved by two steps with a designed chain-based prompting approach. We extract semantic aspects using the \emph{semantic aspects extraction prompt} and then extract aspect-aware reviews based on those aspects using the \emph{semantic aspects-aware review extraction prompt}.

\textbf{Semantic aspect-aware interaction}: if a user $u$'s review contains comments of a specific on a certain aspect of item $i$, then we regard there is an interaction between u and i on this aspect. By identifying the aspects of user reviews on each item, we can form semantic aspect-aware interaction graphs to exploit the semantic aspects in user and item representation learning. 

In previous studies~\cite{wang2019ngcf,He@LightGCN}, the user-item interaction graph is directly built based on the users' historical interactions. Specifically,
let $\mathcal{G}= (\mathcal{W}, \mathcal{E})$ be an undirected graph, where $\mathcal{W}$ denotes the set of nodes and $\mathcal{E}$ is the set of edges. Specifically, $\mathcal{W}$ consists of two types of nodes: users $u_{i} \in \mathcal{U}$ with $i \in \{1,\cdots,N_{u} \}$, items $v_{j} \in \mathcal{V}$ with $j \in \{1,\cdots,N_{v} \}$, and $\mathcal{U}\cup\mathcal{V}=\mathcal{W}$.
By extracting the semantic aspect-aware interactions from reviews using LLM, we separately construct the semantic aspect-specific graphs based on the user-item interactions on various aspects. In this work,
we defined $\mathcal{G}_a = (\mathcal{W}, \mathcal{E}_a)$ as the semantic aspect-based graph for aspect $a$, where $\mathcal{E}_a$ is the set of edges. 

\subsubsection{Embedding Initialization}
In the proposed SAGCN, the vectors of users and items are composed of embeddings from A aspects. Following existing GCN-based recommendation models~\cite{wang2019ngcf, He@LightGCN, Liu2021IMP_GCN}, we initialize the embedding vectors for each aspect. As an example, the ID embeddings of user $u \in \mathcal{U}$ for different aspects are initialized as:
\begin{equation}
    \bm{e_u}^{(0)} = \{\bm{e_u^{(1, 0)}, e_u^{(2, 0)},\cdots,e_u^{(A, 0)}}\},
\end{equation}
where $\bm{e_u^{(0)}} \in \mathds{R}^{d \times A}$ is ID embedding capturing intrinsic characteristics of user $u$; $\bm{e_u^{(a, 0)}} \in \mathds{R}^{{d}}$ is $u$'s representation in aspect $a$. Analogously, $\bm{e_i}^{(0)} = \{\bm{e_i^{(1, 0)}, e_i^{(2, 0)},\cdots,e_i^{(A, 0)}}\}$ is defined as the ID embeddings of item $i \in \mathcal{I}$. 
In addition, it is worth noting that our model randomly and separately initializes the embeddings of the user IDs and item IDs for various aspects.

\subsubsection{Embedding Propagation}
To exploit the high-order connectives on each aspect to learn user and item representations,
we propagate embeddings across various semantic aspect-based graphs. Take the semantic aspect-specific graph $\mathcal{G}_a$ as an example, after propagating embeddings through $k$ graph convolution layers,
we define $\bm{e_{u}^{(a, k)}}$ and $\bm{e_{i}^{(a, k)}}$ to denote the embedding of user $u$ and item $i$, which are propagated from $k-$th order neighboring nodes, where $k \geq 0$, the embedding propagation is formulated as:
\begin{equation}
\begin{aligned}
\bm{e_{u}^{(a, k+1)}} =  \sum_{i \in \mathcal{N}_{u}^a}\frac{1}{\sqrt{ | \mathcal{N}_{u}^a | }\sqrt{ | \mathcal{N}_{i}^a |}}\bm{e_{i}^{(a, k)}}, \\
\bm{e_{i}^{(a, k+1)}} =  \sum_{u \in \mathcal{N}_{i}^a}\frac{1}{\sqrt{ | \mathcal{N}_{i}^a | }\sqrt{ | \mathcal{N}_{u}^a |}}\bm{e_{u}^{(a, k)}},
\end{aligned}
\end{equation}
where $\bm{e_{i}^{(a, k+1)}}$ and $\bm{e_{u}^{(a, k+1)}}$ can be considered as embeddings propagated from the $(k+1)$-th order neighboring nodes. $\frac{1}{\sqrt{ | \mathcal{N}_u^{a} | }\sqrt{ | \mathcal{N}_i^{a} |}}$ is symmetric normalization terms; $\mathcal{N}_u^{a}$ denotes the set of items that interact with user $u$, and $\mathcal{N}_i^{a}$ denotes the set of users that interact with item $i$.

\subsubsection{Embedding Combination}
After the $K$-layers graph convolution, the embeddings of user $u$ and item $i$ of aspect $a$ are aggregated from all layers:
\begin{equation}
\begin{aligned}
\bm{e_{u}^{(a)}}=\sum_{k=0}^{K}\bm{e_{u}^{(a, k)}},
\bm{e_{i}^{(a)}}=\sum_{k=0}^{K}\bm{e_{i}^{(a, k)}}.
\end{aligned}
\end{equation}
The final representations of user $u$ and item $i$ are a concatenation of their embeddings learning in multiple semantic aspect-based graphs:
\begin{equation}
\begin{aligned}
\bm {e_u} = \bm{e_{u}^{(0)}}||\cdots||\bm{e_{u}^{(a)}}||\cdots||\bm{e_{u}^{(A)}},
\bm {e_i} = \bm{e_{i}^{(0)}}||\cdots||\bm{e_{i}^{(a)}}||\cdots||\bm{e_{i}^{(A)}},
\end{aligned}
\end{equation}
where $\bm{e_{u}^{(a)}}$ and $\bm{e_{i}^{(a)}}$ denote the embeddings of users and items by performing graph convolution on graph $\mathcal{G}_a$, respectively. 

\subsubsection{Model Training}
After learning the representation for both users and items, the preference of a given user $u$ for a target item $i$ is estimated as follows:
\begin{equation}
    \hat{r}_{ui} = \bm{e_{u}}^T\bm{e_{i}}.
\end{equation}

In this work, we aim to recommend a list of items to the target users by computing their preference score. For the optimization, we utilize the pairwise learning method, which is similar to the approach employed in previous works~\cite{He@LightGCN,Liu2021IMP_GCN}. To train our SAGCN model, we construct a triplet of $\{u, i^+, i^-\}$, comprising a positive item (an observed interaction between $u$ and $i^+$) and a negative item (an unobserved interaction between $u$ and $i^-$). The objective function is formulated as follows:
\begin{equation}
\mathop{\arg\min} \sum_{(\mathbf{u}, \mathbf{i}^+,\mathbf{i}^-)\in{\mathcal{O}}} -\ln\phi(\hat{r}_{ui^+} - \hat{r}_{ui^-}) + \lambda\left\|\Theta\right\|^2_2
\end{equation}
where $\mathcal{O}$ represent the training set, defined as $\mathcal{O}=\{(u, i^+, i^-)|(u,i^+)\in\mathcal{R^+}, (u,i^-) \in\mathcal{R^-}\}$. Here, $\mathcal{R^+}$ indicates the observed interactions in the training dataset, while $\mathcal{R^-}$ denotes a set of sampled unobserved interactions. Additionally, $\lambda$ represents the regularization weight, and $\Theta$ stands for the model parameters. We employ $L_2$ regularization to mitigate overfitting. To optimize our model and update its parameters, we utilize the mini-batch Adam technique, as described in~\cite{kingma2014adam}. 

\section{Experiments}
\label{sec:experiments}
\subsection{Experimental setup}

\subsubsection{Data Description.}
To evaluate the effectiveness of our approach, we conducted experiments on Amazon Product Datasets and Goodreads Review Datasets.

\begin{table}[t]
	\centering
	\caption{ Statistics of the Experimental Datasets.}
	\label{tab:data}
	\begin{tabular}{c|c|c|c|c}
		\hline 
		Dataset&\#user&\#item&\#interactions &sparsity \\ \hline  
		Office & 4,905 & 2,420 & 53,258  & 99.55\% \\ \hline 
        {Goodreads} & 4,545 & 5,274 & 53,458  & 99.78\% \\ \hline 
		  Baby & 19,445 & 7,050 &160,792  & 99.88\% \\ \hline
            Clothing & 39,387 & 23,033 & 278,677  & 99.97\% \\ 
		\hline 
	\end{tabular}
	\vspace{0pt}
\end{table}

~\textbf{Amazon Product Datasets}~\footnote{http://jmcauley.ucsd.edu/data/amazon}. 
These datasets stem from the extensive product review system of Amazon. It includes user-product interactions, reviews, and metadata. In our experiments, we focused solely on user-product interactions and user reviews. Three product categories from the 5-core version of these datasets are selected for our evaluation: Office Products (Office), Baby, and Clothing. This indicates that both users and items have a minimum of 5 interactions.

{\textbf{Goodreads Review Datasets}~\footnote{https://www.goodreads.com/}. The datasets stem from the product review of GoodReads, which is the largest site for readers and book recommendations. It collected user-book interactions, reviews, and book metadata with anonymized user IDs and book IDs. In our experiments, we only use user-book interactions and review text.}

In line with previous studies on conventional recommendation methods~\cite{He@LightGCN,Liu2021IMP_GCN}, we randomly split each user's interaction into training and testing sets with a ratio of 80:20 for each dataset. Within the training set, we randomly select 10\% of interactions to establish the validation set for hyper-parameter optimization. The pairwise learning strategy is adopted for parameter optimization. Specifically, the observed user-item interactions in the training set were treated as positive instances. For each positive instance, we paired it with a random sampling of negative instances that the user had not interacted with yet. For a fair comparison, we consistently adopt the negative sampling strategy across our methods and competitors. To further validate the effectiveness of our approach, we compare our proposed SAGCN with a pure LLM-based method, and the details of test dataset construction for the LLM-based method can be seen in Section~\ref{section:LLM_comparison}. 

\subsubsection{Compared Baselines.}
To demonstrate the effectiveness of our approach, we compared our SAGCN with several state-of-the-art models, which include CF-based recommendation methods without side information ($i.e.$, NeuMF, GCMC, LightGCN, DGCF, NCL) and multimodal recommendation methods ($i.e.$, MMGCN, GRCN, LATTICE, BM3). For fair comparisons, we only adopt reviews as the side information for MMGCN, GRCN, LATTICE, and BM3. In addition, we devised a LLM-based recommendation method for comparison, the details are provided in Section~\ref{section:LLM_comparison}.

\begin{itemize}[leftmargin=*]
\item \textbf{NeuMF~\cite{he2017neural}}: This is a state-of-the-art neural collaborative filtering method that present a general framework for collaborative filtering based on neural networks. It innovatively introduces multiple hidden layers to capture the non-linear interactions between the latent features of users and items. 

\item \textbf{GCMC~\cite{berg2019gcmc}}: It is a graph-based auto-encoder framework for matrix completion. This method leverages the GCN technique~\cite{kipf2017gcn} on the user-item interaction graph and solely uses one convolutional layer to exploit direct connections between the nodes of users and items. 

\item \textbf{NGCF~\cite{wang2019ngcf}}: This is the very first GCN-based recommendation model that incorporates high-order connectivity of user-item interactions. It encodes the collaborative signal from high-order neighbors by performing embedding propagation on the user-item bipartite graph.

\item \textbf{LightGCN~\cite{He@LightGCN}}: It is a state-of-the-art graph collaborative filtering method. This model can be considered a simplified version of NGCF~\cite{wang2019ngcf}, where certain components such as the feature transformation and non-linear activation module are removed. 

\item \textbf{IMP-GCN~\cite{Liu2021IMP_GCN}}: This method performs high-order graph convolution within sub-graphs, each of which is constructed based on users with shared interests and associated items, thus avoiding the negative information passed from high-order neighbors. 

\item \textbf{NCL~\cite{lin2022NCL}}: This is a model-agnostic contrastive learning framework, which explicitly incorporates the potential neighbors into contrastive pairs. It aims to capture the correlation between a node and its prototype, thereby refining the neural graph collaborative filtering techniques. 

\item \textbf{LATTICE~\cite{LATTICE2021TMM}}: This method uses side information to construct an item-item graph via a modality-aware structure learning layer and perform graph convolution operations to aggregate information from high-order neighbors to enhance the item representations.

\item \textbf{MMGCN~\cite{wei2019mm}}: The model learns user preferences and item characteristics by propagating modality embeddings on the user-item interaction graph. 

\item \textbf{GRCN~\cite{Wei2019GRCN}}: This GCN-based multimedia recommendation model designs a graph refining layer to adjust the structure of the interaction graph adaptively based on the status of model training for discovering and pruning potential false-positive edges.

\item \textbf{BM3~\cite{BM32020Arxiv}}: This self-supervised framework uses a latent representation dropout mechanism to generate the target view of a user or an item. Moreover, it is trained without negative samples by jointly optimizing three multi-modal objectives.

\item {\textbf{DeepCoNN ~\cite{DeepCoNN2017WSDM}}: This model jointly learns item properties and user behavior from review text by using Convolutional Neural Networks (CNNs) to encode the textual data.}

\item {\textbf{NARRE~\cite{chen2018neural}}: The model uses two parallel CNNs to learn the latent representation of reviews. Then, an attention mechanism is designed to estimate the contributions of different reviews for the item and user.}

\item {\textbf{RGCL~\cite{Shuai2022SIGIR}}:
This model constructs a review-aware user-item graph with feature-enhanced edges derived from reviews. Each edge feature incorporates both the user-item rating and the semantic information extracted from the corresponding review.}

\item {\textbf{LightGCN$_{LDA}$~\cite{blei2003latent}}: To validate the ability of LLMs in understanding reviews, we designed a baseline that combines LDA with LightGCN. LDA extracts topics from the dataset, and we calculate the similarity between each review and these topics. Using these similarities, we construct topic-specific graphs, where LightGCN learns user and item representations through graph convolutions.}

\item {\textbf{LlamaRec~\cite{yue2023llamarec}}: The model is a two-stage ranking-based recommendation. This method first retrieves a series of candidate items by small-scale recommenders learning based on the user interaction history. Then, the model adopts an LLM-based ranking model to reranking the order of candidate items.}

\item {\textbf{RLMRec~\cite{ren2024representation}}: The model designs an LLM-based generation framework that generates user and item profiles based on interaction data and textual information, such as product descriptions, user reviews, and product titles. Then, these generated profiles are encoded as semantic information of LLMs and incorporated into the CF-based recommendation system.}

\end{itemize}

\subsubsection{\textbf{Evaluation Metrics}}
Two widely-used evaluation metrics are employed for evaluating top-$K$ recommendations: \emph{Recall} and \emph{Normalized Discounted Cumulative Gain} (NDCG)~\cite{he2015trirank}. 
For conventional recommendation methods, the recommendation accuracy is calculated for each metric based on the top 10 and 20 results. The reported results are the average values across all testing users. In contrast, for the LLM-based method we have devised for comparison, we calculate the recommendation accuracy based on the top 1, 3, 5, and 7 results. In addition, the reported results are the average values across the sampled users. Note that we sample several users for evaluation due to the high resource and time consumption of LLMs. 

\subsubsection{\textbf{Parameter Settings}}
The PyTorch framework~\footnote{https://pytorch.org.} is adopted to implement the proposed model. We employed Vicuna-13B ~\footnote{https://huggingface.co/lmsys/vicuna-13b-v1.5} to analyze the user reviews. 
It has been fine-tuned from LLaMA using supervised instruction-based fine-tuning and exhibits strong performance in sentiment analysis. In our experiments, all hyperparameters are carefully tuned. The user and item embedding size for all competitor methods is set to 64 across all datasets. The number of semantic aspects $N$ is defined as 8 across all four datasets. The embedding size associated with each aspect is also set to 64 within our proposed SAGCN. The mini-batch size is fixed to 1024. The learning rate for the optimizer is searched from $\{1e^{-5},1e^{-4},\cdots,1e^{1}\}$, and the model weight decay is searched in the range $\{1e^{-5}, 1e^{-4},\cdots, 1e^{-1}\}$. 
Besides, the early stopping strategy is implemented to enhance training efficiency.
Specifically, the training process will stop if recall@10 does not increase for 20 successive epochs. 

\subsection{{\textbf{Performance Comparison With Baseline Methods}}}
\begin{table*}[t]
	\vspace{0pt}
	\caption{Performance of our method and the competitors over four datasets.} 
	\centering
 \resizebox{1.0\textwidth}{!}{
\begin{tabular}{l|cccc|cccc|cccc|cccc}
\hline
Datasets & \multicolumn{4}{c|}{Office}                                         & \multicolumn{4}{c|}{Baby}                                          & \multicolumn{4}{c|}{Clothing}                                     & \multicolumn{4}{c}{Goodreads}                                         \\ \cline{2-17} 
Metrics  & R@10            & N@10           & R@20            & N@20           & R@10           & N@10           & R@20            & N@20           & R@10           & N@10           & R@20           & N@20           & R@10            & N@10            & R@20            & N@20            \\ \hline
NeuMF    & 5.14            & 3.89           & 8.12            & 5.21           & 3.11           & 2.11           & 4.85            & 2.69           & 0.94           & 0.54           & 1.50           & 0.71           & 9.24            & 7.45            & 14.63           & 9.02            \\
GCMC     & 6.72            & 5.27           & 10.27           & 6.79           & 4.55           & 2.99           & 7.24            & 3.89           & 3.17           & 1.86           & 4.86           & 2.35           & 16.32           & 10.58           & 22.32           & 12.32           \\
LightGCN & 9.87            & 6.04           & 14.47           & 7.43           & 5.94           & 3.30           & 9.25            & 4.20           & 4.45           & 2.43           & 6.44           & 2.95           & 16.99           & 11.02           & 23.35           & 13.01           \\
DGCF     & 9.95            & 6.25           & 14.37           & 7.52           & 5.90           & 3.27           & 9.20            & 4.14           & 4.67           & 2.68           & 6.91           & 3.32           & 16.89           & 11.22           & 21.76           & 13.12           \\
IMP-GCN  & 10.11           & 6.36           & 14.47           & 7.71           & 6.24           & 3.49           & \textbf{9.56}   & 4.38           & 4.80           & 2.76           & 7.11           & 3.40           & 18.05           & 12.07           & 23.79           & 14.12           \\
NCL      & 10.07           & 6.30           & 14.40           & 7.65           & 6.15           & 3.42           & 9.45            & 4.30           & 4.76           & 2.74           & 7.10           & 3.37           & 17.69           & 11.64           & 23.55           & 13.68           \\ \hline
MMGCN    & 5.74            & 3.42           & 9.39            & 4.54           & 3.95           & 2.17           & 6.46            & 2.85           & 2.42           & 1.29           & 3.76           & 1.64           & 11.24           & 7.85            & 15.33           & 9.48            \\
GRCN     & \textbf{10.38}  & 6.34           & \textbf{15.33}  & \textbf{7.81}  & 5.57           & 3.03           & 8.49            & 3.83           & 4.47           & 2.35           & 6.70           & 2.94           & 16.62           & 10.79           & 22.64           & 12.67           \\
LATTICE  & 10.00           & 6.09           & 14.99           & 7.57           & 6.06           & 3.40           & 9.29            & 4.27           & 5.03           & 2.79           & 7.28           & 3.37           & 17.92           & 11.35           & 23.60           & 13.27           \\
BM3      & 9.80            & 6.09           & 14.02           & 7.36           & \textbf{6.44}  & \textbf{3.65}  & 9.52            & \textbf{4.48}  & \textbf{5.28}  & \textbf{2.94}  & \textbf{7.75}  & \textbf{3.58}  & \textbf{18.28}  & \textbf{12.12}  & \textbf{24.68}  & \textbf{14.15}  \\ \hline
{DeepCoNN} & 5.32            & 4.01           & 8.35            & 5.33           & 3.20           & 2.02           & 5.05            & 2.71           & 1.89           & 1.01           & 2.98           & 1.35           & 11.89           & 7.22            & 5.66            & 8.44            \\
{NARRE}    & 6.12            & 4.78           & 9.41            & 6.15           & 4.02           & 2.32           & 6.14            & 2.98           & 2.37           & 1.32           & 3.62           & 1.69           & 12.79           & 9.32            & 16.84           & 10.89           \\
{RGCL}     & 7.89            & 5.69           & 12.40           & 7.02           & 5.22           & 2.54           & 8.20            & 3.19           & 3.32           & 1.93           & 5.22           & 2.41           & 16.55           & 10.43           & 21.79           & 12.19           \\
{LightGCN$_{LDA}$} & 10.12           & 6.24           & 15.12           & 7.67           & 5.98           & 3.44           & 9.41            & 4.29           & 5.17           & 2.88           & 7.50           & 3.44           & 16.73           & 10.87           & 22.07           & 12.70           \\ \hline
{LlamaRec} & 9.89            & 5.83           & 14.44           & 7.20           & 5.62           & 3.11           & 8.83            & 3.85           & 3.72           & 2.17           & 5.45           & 2.68           & 16.67           & 10.82           & 21.80           & 12.66           \\
{RLMRec}   & 10.1            & \textbf{6.35}  & 14.87           & 7.78           & 5.84           & 3.37           & 8.91            & 4.30           & 4.48           & 2.60           & 6.65           & 3.08           & 16.72           & 11.02           & 22.02           & 12.71           \\ \hline
{SAGCN}    & \textbf{11.71*} & \textbf{7.34*} & \textbf{16.71*} & \textbf{8.84*} & \textbf{7.35*} & \textbf{4.23*} & \textbf{10.56*} & \textbf{5.09*} & \textbf{6.07*} & \textbf{3.58*} & \textbf{8.44*} & \textbf{4.20*} & \textbf{19.40*} & \textbf{13.15*} & \textbf{26.17*} & \textbf{15.14*} \\ \hline
{Imporv.}  & 12.81\%         & 15.59\%        & 9.00\%          & 13.19\%        & 14.13\%        & 15.89\%        & 10.46\%         & 13.61\%        & 14.96\%        & 21.77\%        & 8.90\%         & 17.32\%        & 6.15\%          & 8.50\%          & 6.03\%          & 7.03\%          \\ \hline
\end{tabular}}
	\begin{tablenotes}
		\footnotesize
		\item The symbol * denotes that the improvement is significant with $p-value < 0.05$ based on a two-tailed paired t-test. Notice that the values are reported by percentage with '\%' omitted.
	\end{tablenotes}
	\label{tab:results}
	\vspace{0pt}
\end{table*}

In this section, we compared the performance of our proposed SAGCN with all the adopted competitors (conventional recommendation models). The experimental results of our method and all compared methods over four test datasets in terms of Recall@10, Recall@20, NDCG@10, and NDCG@20 are reported in Table~\ref{tab:results}. Note that the best and second-best results are highlighted in bold. From the results, we have the following observations.

We first focus on the performance of methods in the first block, which only uses the user-item interaction information. NeuMF is a deep learning-based method, and it shows relatively poor performance. GCMC performs better than NeuMF, highlighting the benefits of leveraging graph structure. However, GCMC underperforms LightGCN because it not explicitly leverage high-order connectivities between users and items. 
This showcases the strength of GCN and the importance of incorporating high-order information into representation learning. Moreover, DGCF performs better than LightGCN by leveraging the disentangled representation technique to model users' diverse intents.
Both IMP-GCN and NCL consistently outperform all the above GCN-based methods over all the datasets. This is mainly due to their ability to filter out noisy information from neighboring nodes effectively. 
In particular, IMP-GCN's performance superiority over LightGCN and DGCF demonstrates the importance of distinguishing nodes in high-order neighbors in the graph convolution operation. 

The second block is the GCN-based methods utilizing user reviews as side information. MMGCN is the first method combining side information with GCN, it exploits the high-order information of various modalities. However, it falls short of GRCN in terms of effectiveness. The reason is that the GRCN could adaptively adjusting the false-positive edges of the interaction graph to discharge the ability of GCN. LATTICE outperforms GRCN on Baby and Clothing by constructing item-item graphs based on item similarity in the side information. BM3 consistently performs better than all general CF baselines over Baby and Clothing. It should be credited to the joint optimization of multiple objectives to learn the representations of users and items.

{The third block illustrates that review-based methods primarily use user textual information as a key semantic feature. DeepCoNN, which employs Convolutional Neural Networks, designs two coupled networks to learn user and item representations from reviews, outperforming NeuMF due to its use of semantic information. NARRE enhances recommendations by using attention mechanisms to weigh the significance of different reviews, showing particular improvement on the sparse Clothing dataset. RGCL further incorporates review data by building a review-aware user-item graph, outperforming all three review-based methods. To validate the ability of LLMs in understanding reviews, we designed LightGCN$_{LDA}$, which combines LDA with LightGCN. Similar to our approach, it uses topic-level information to construct user-item graphs and then learns user and item representations through GCN techniques. As a result, it outperforms all review-based models across all datasets.}

{The fourth block consists of two LLM-based methods: LlamaRec and RLMRec. LlamaRec is a two-stage framework that first retrieves candidate items and then ranks them using an LLM. It outperforms NeuMF due to the advanced capabilities of LLMs in understanding complex patterns and relationships within the data. RLMRec, on the other hand, outperforms LlamaRec by leveraging a more comprehensive data approach and generating richer user and item profiles. While LlamaRec focuses solely on user interaction data, RLMRec combines interaction data with textual information such as product descriptions, user reviews, and product titles. This enables RLMRec to create detailed, contextually enriched profiles, offering a deeper understanding of user preferences and item characteristics.
}

{SAGCN consistently provides the best performance across all datasets. In particular, the improvements of SAGCN over the strongest competitors $w.r.t$ NDCG@10 are 15.59\%, 15.89\%, 21.77\%, and 8.50\% in Office, Baby, Clothing, and Goodreads, respectively. We credit this to the combined effects of the following three aspects. Firstly, we accurately extract structured semantic aspect-aware reviews using LLM to understand user behaviors at a finer level, which also avoids the noise in raw reviews. Our proposed method outperforms LightGCN$_{LDA}$, highlighting the superior understanding capabilities of LLMs. Additionally, the semantic aspects generated by LLMs offer better interpretability compared to the latent topics extracted by LDA.} Secondly, the SAGCN performs graph convolution on various semantic-aspect-based graphs, which can alleviate the over-smoothing problem~\cite{Liu2021IMP_GCN} and avoid negative information from unreliable neighbor nodes~\cite{Wei2019GRCN}. Thirdly, we extract fine-grained user behaviors for each semantic aspect to bridge the modality gap between interaction and textual features. Different with existing methods that integrate textual features with user-item interaction data for representation learning, our approach constructs structured user-item interactions from user reviews via LLM, which are associated with various semantic aspects. These structured interactions are then leveraged for the representation learning of users and items.

\subsection{{\textbf{Performance Comparison with Pure LLM}}}
\label{section:LLM_comparison}

To further validate the effectiveness of our approach, we implemented a method named \textbf{LLM$^*$}, built upon LLM for comparative analysis. However, due to the high resource and time consumption of LLM, we evaluated the method by randomly selecting 200 users from each dataset. 
To construct the testing data, we randomly selected one positive item from the history items of the original testing data. We then sampled 9 negative items not interacted with by the user as the testing set for each user.
Next, we converted the historical data and items in the testing set for each user into textual format using a prompt template. Finally, we fed the converted text into LLM to generate scores for the items in the testing set. These items are then ranked according to the scores generated by the LLM. Our prompt template is as follows:
\begin{quote}
\emph{I want you to rate every candidate product's historical record of purchased habits. You are encouraged to learn his preferences from the historical records he has purchased. Here are The historical interactions of a user include: \{ history \}. Now, how will the user rate these candidate products? (1 being lowest and 5 being highest) \{ candidates \}. Importantly, the interacted items should have been excluded from the rating. Finally, Only output rating item list, which template is: 1. Swingline GBC UltraClear Thermal Laminating Pouches, Menu Size, 3 Mil, 25 Pack (item id: B00006IA2K) - Rating: 4.0 stars} 
\end{quote}
Where \emph{\{history\}} consists of item title and user rating score; \emph{\{candidates\}} is composed of item titles and item id. Besides, we also use the \textbf{Random} method as a comparison method, which randomly ranks the items in the testing set.

As illustrated in Tables \ref{tab:office_LLM_results} and~\ref{tab:clothing_LLM_results}, the performance of the Random method is inferior as it randomly recommends the items to users. The LLM$^{*}$ method achieve better performance than Random. This is because it has the ability to recognize patterns and correlations in data. This capability allows it to predict user preferences and make more informed recommendations based on observed patterns in user behavior or item characteristics. However, its performance does not significantly exceed that of Random, especially in terms of Recall@1 and NDCG@1. This can be attributed to the LLM not having been trained or fine-tuned for the specific recommendation task.
Experimental results demonstrate the effectiveness of our proposed approach, which integrates LLM with conventional recommendation techniques. A remarkable advantage of SAGCN is its aspect-aware graph, which allows for a finer understanding of complex interactions between users and items, further augmented by the sentiment analytical power of LLM. 

\begin{table*}[t]
	\vspace{0pt}
	\caption{Performance Comparison of Random, LLM* and our method on Office.} 
	\centering
\begin{tabular}{c|llllllcl}
\hline
Datasets                 & \multicolumn{8}{c}{Office}                                                                                                                                                                                         \\ \hline
\multirow{2}{*}{Metrics} & \multicolumn{2}{c|}{k=1}                                 & \multicolumn{2}{c|}{k=3}                                 & \multicolumn{2}{c|}{k=5}                                 & \multicolumn{2}{c}{k=7}           \\ \cline{2-9} 
                         & \multicolumn{1}{c}{Recall} & \multicolumn{1}{c|}{NDCG}   & \multicolumn{1}{c}{Recall} & \multicolumn{1}{c|}{NDCG}   & \multicolumn{1}{c}{Recall} & \multicolumn{1}{c|}{NDCG}   & Recall & \multicolumn{1}{c}{NDCG} \\ \hline
Random                   & 0.0835                     & \multicolumn{1}{l|}{0.0835} & 0.2926                     & \multicolumn{1}{l|}{0.2012} & 0.5261                     & \multicolumn{1}{l|}{0.2962} & 0.7203 & 0.3632                   \\ \hline
LLM$^{*}$                & 0.1086                     & \multicolumn{1}{l|}{0.1086} & 0.5803                     & \multicolumn{1}{l|}{0.3903} & 0.7660                     & \multicolumn{1}{l|}{0.4657} & 0.9122 & 0.5163                   \\ \hline
SAGCN                    & 0.4880                     & \multicolumn{1}{l|}{0.4880} & 0.7470                     & \multicolumn{1}{l|}{0.6371} & 0.8763                     & \multicolumn{1}{l|}{0.6838} & 0.9669 & 0.7289                   \\ \hline\hline
\end{tabular}
	\label{tab:office_LLM_results}
	\vspace{0pt}
\end{table*}

\begin{table*}[t]
	\vspace{0pt}
	\caption{Performance Comparison of Random, LLM* and our method on Clothing.} 
	\centering
\begin{tabular}{c|llllllcl}
\hline
Datasets                 & \multicolumn{8}{c}{Clothing}                                                                                                                                                                                        \\ \hline
\multirow{2}{*}{Metrics} & \multicolumn{2}{c|}{k=1}                                 & \multicolumn{2}{c|}{k=3}                                 & \multicolumn{2}{c|}{k=5}                                 & \multicolumn{2}{c}{k=7}           \\ \cline{2-9} 
                         & \multicolumn{1}{c}{Recall} & \multicolumn{1}{c|}{NDCG}   & \multicolumn{1}{c}{Recall} & \multicolumn{1}{c|}{NDCG}   & \multicolumn{1}{c}{Recall} & \multicolumn{1}{c|}{NDCG}   & Recall & \multicolumn{1}{c}{NDCG} \\ \hline
Random                   & 0.0841                     & \multicolumn{1}{l|}{0.0841} & 0.2848                     & \multicolumn{1}{l|}{0.2208} & 0.4474                     & \multicolumn{1}{l|}{0.2867} & 0.6186 & 0.3458                   \\ \hline
LLM$^{*}$                & 0.0912                     & \multicolumn{1}{l|}{0.0912} & 0.4860                     & \multicolumn{1}{l|}{0.2997} & 0.7692                     & \multicolumn{1}{l|}{0.4170} & 0.8913 & 0.4669                   \\ \hline
SAGCN                    & 0.4823                     & \multicolumn{1}{l|}{0.4823} & 0.7239                     & \multicolumn{1}{l|}{0.6176} & 0.8273                     & \multicolumn{1}{l|}{0.6603} & 0.9245 & 0.6908                   \\ \hline\hline
\end{tabular}
	\label{tab:clothing_LLM_results}
	\vspace{0pt}
\end{table*}

\subsection{\textbf{Effect of Layer Numbers}}
\begin{figure*}[t]
	\centering
         \subfloat[Recall on Office]{\includegraphics[width=0.25\linewidth]{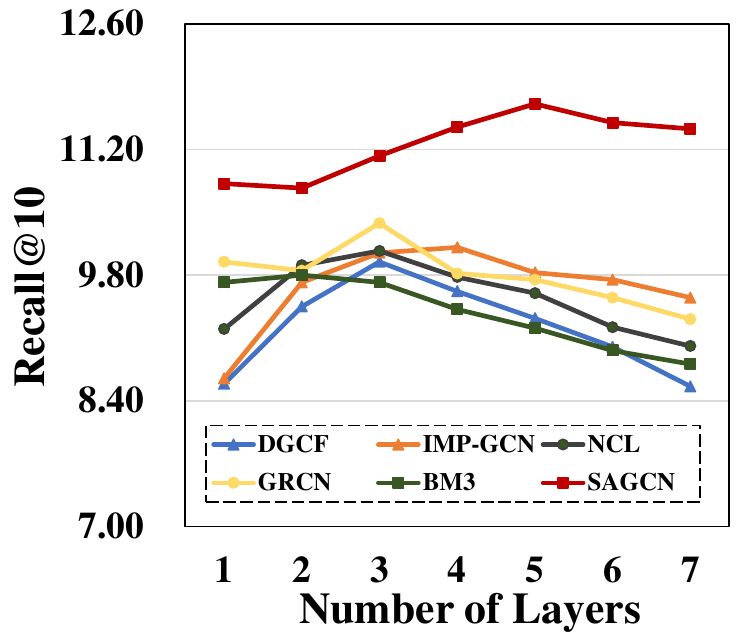}}
         \subfloat[Recall on Clothing]{\includegraphics[width=0.25\linewidth]{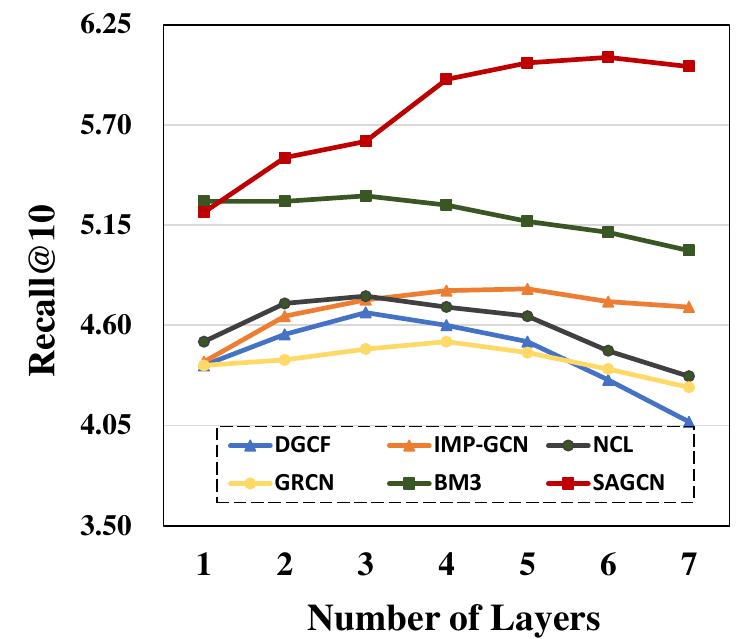}}
         \subfloat[Recall on Baby]{\includegraphics[width=0.25\linewidth]{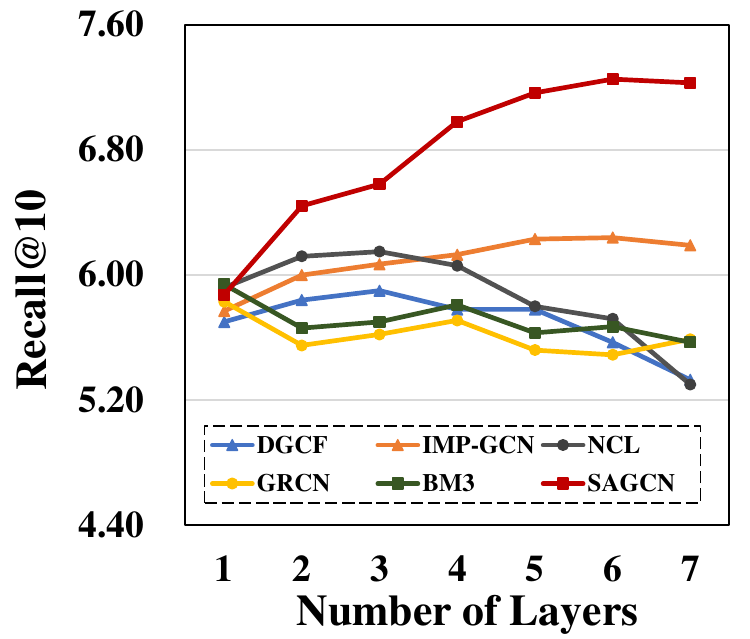}}
         \subfloat[{Recall on Gooodreads}]{\includegraphics[width=0.25\linewidth]{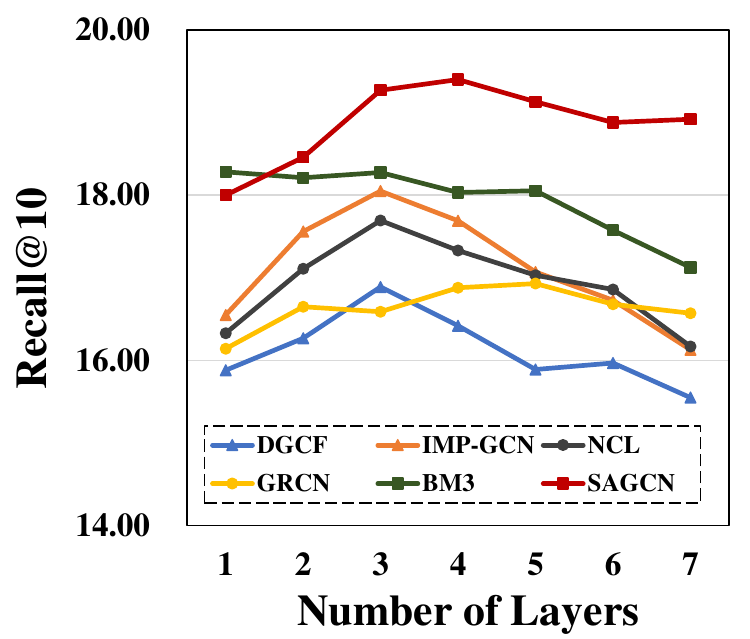}}
         \newline
         \subfloat[NDCG on Office]{\includegraphics[width=0.25\linewidth]{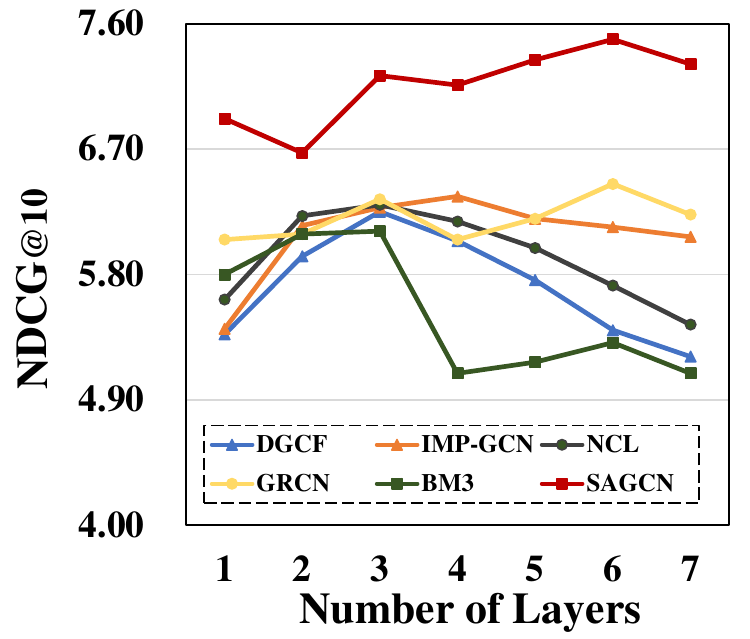}}
         \subfloat[NDCG on Clothing]{\includegraphics[width=0.25\linewidth]{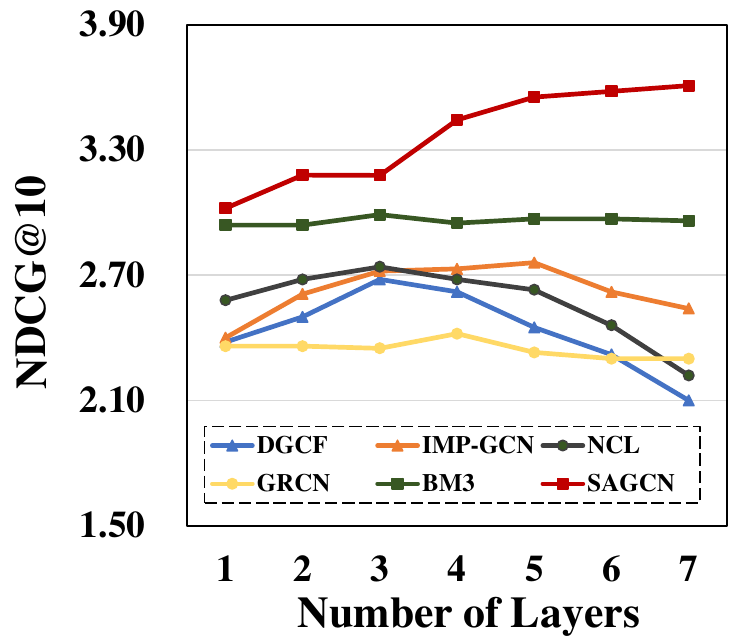}}
         \subfloat[NDCG on Baby]{\includegraphics[width=0.25\linewidth]{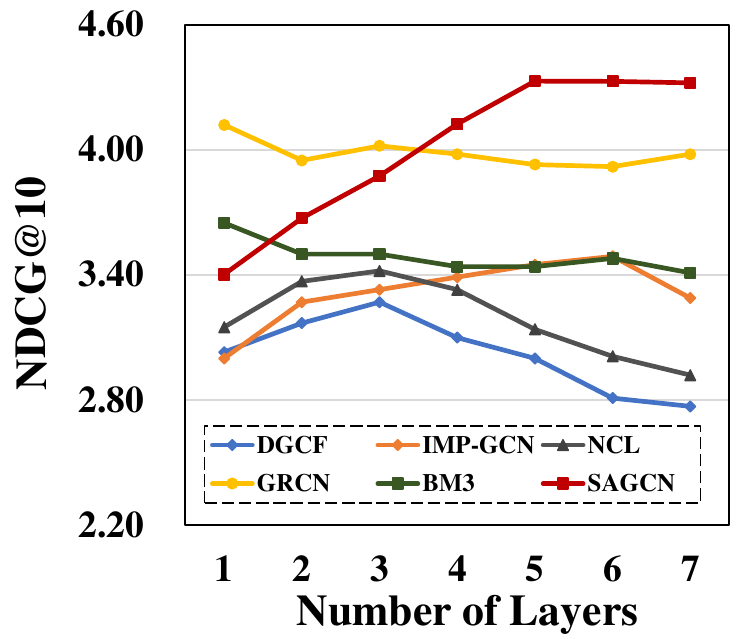}}
         \subfloat[{NDCG on Gooodreads}]{\includegraphics[width=0.25\linewidth]{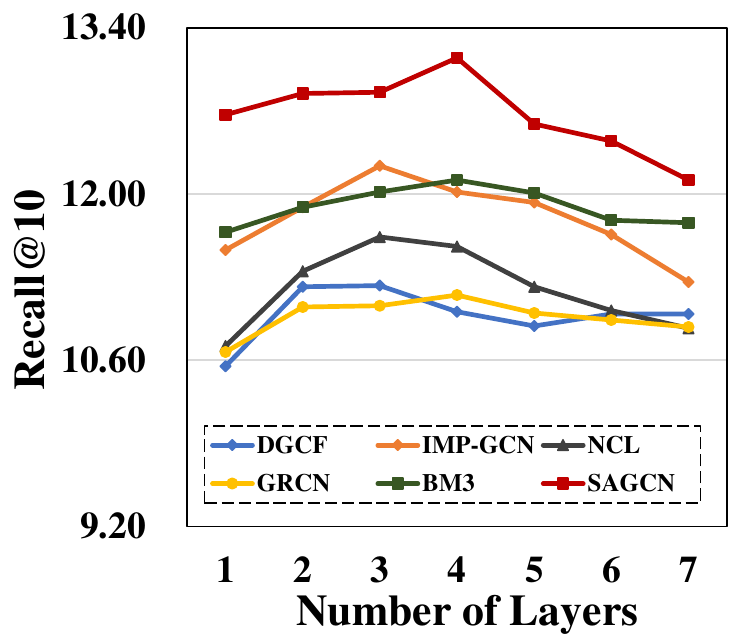}}
         \newline
        \hspace{0mm}
	\caption{{Performance Comparison between SAGCN and competitors at different layers on Office, Clothing, Baby, and Goodreads. Notice that the values are reported by percentage with '\%' omitted.}}
\label{fig:GCNlayer}
\end{figure*}

In this section, we evaluated the performance of our model by varying the number of layers in the graph convolution process. Our primary goal was to examine whether our model can mitigate the over-smoothing problem often encountered in graph convolutions.
To investigate the effectiveness of our model, we iteratively increased the number of convolution layers from 1 to 7. We then compared its performance against four GCN-based methods: LightGCN, DGCF, IMP-GCN, GRCN, and BM3. The comparative outcomes are depicted in Figure~\ref{fig:GCNlayer}. 

From the results, all three models (DGCF, NCL, and BM3) achieve their peak performance when stacking 2 or 3 layers across two datasets. Beyond this, they will cause dramatic performance degradation with increasing layers. These results indicate that these three models suffer from the over-smoothing problem and inadvertently introduce noise information in the deep structure. GRCN achieves the peak performance when stacking 4 layers. This is attributed to discovering and pruning the potential false-positive edges, which avoid the noisy information from unreliable neighboring nodes. IMP-GCN can achieve robust performance with a deeper structure. This resilience can be credited to its effectiveness in filtering out noisy information from high-order neighbors. Our proposed SAGCN consistently outperforms all competitors when more than 2 layers are stacked, and it achieves peak performance when stacking 5 or 6 layers. This indicates that SAGCN can alleviate the over-smoothing problem by performing graph convolution over multiple semantic aspect-based graphs. 

\subsection{{Handling of Noisy Neighboring Nodes}}
\begin{figure}[t]
	\centering	
        \includegraphics[width=1\linewidth]{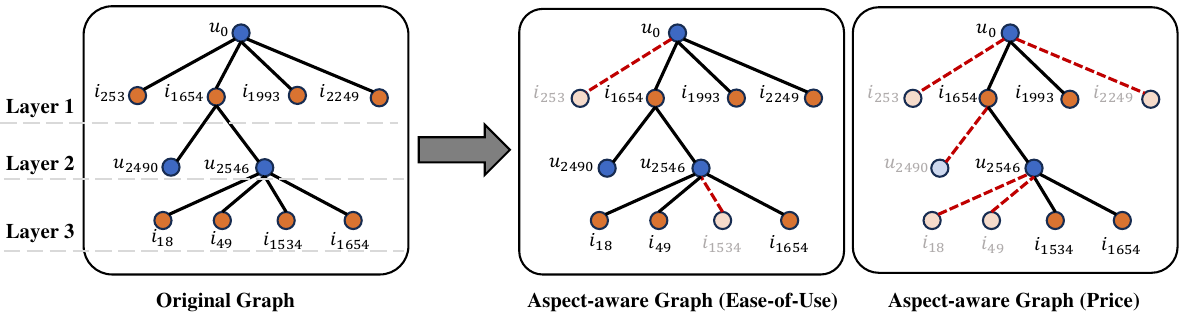}
	\caption{{Illustration of Aspect-Aware Graph Construction for Reducing Noisy Neighboring Nodes in Graph Convolutions. User $u_0$ is the target node, and we selected some of its first, second, and third-order neighboring nodes to show their connectives in the original graph and aspect-aware graphs.}}
	\vspace{-0pt}
\label{fig:noisynodes}
\end{figure}

{We conducted a case study to illustrate how our approach handles noisy neighboring nodes. As shown in Fig.~\ref{fig:noisynodes}, we selected user $u_0$ as the target node and chose some of its first, second, and third-order neighboring nodes to construct a graph displaying their connections. In the original graph, when performing graph convolutions, information from all neighboring nodes is passed to the target node $u_0$ through the same graph structure. In contrast, our approach, which leverages LLMs to analyze the semantic aspects mentioned in each review, constructs aspect-aware graphs based on these results. This means that certain neighboring nodes may only appear in specific aspect-aware graphs. For instance, in user $u_{2546}$'s review of $i_{1534}$, only the aspect of Price was mentioned, while Ease-of-Use was not. As a result, when constructing the graph, node $i_{1534}$ only appears in the Price aspect-aware graph and passes information to the target node through this specific graph structure.}

{By constructing aspect-aware graphs based on specific semantic aspects identified in user reviews, our method ensures that only the most relevant information is passed from neighboring nodes to the target node. In traditional graph convolution methods, all neighboring nodes contribute information equally, even if their relevance to the target node is minimal or unrelated, which introduces noise and dilutes the signal from more pertinent nodes. With our approach, neighboring nodes that are not relevant to a specific aspect (e.g., Ease-of-Use or Price) are excluded from the corresponding aspect-aware graph. This selective filtering reduces the contribution of irrelevant or less important nodes, allowing the target node to receive more focused and meaningful information. As a result, the quality of the information passed to the target node improves, leading to more accurate representations and better overall model performance.}

\subsection{{\textbf{Effect of Embedding Dimensions}}}
\begin{figure*}[t]
	\centering
         \subfloat[Recall on Office]{\includegraphics[width=0.25\linewidth]{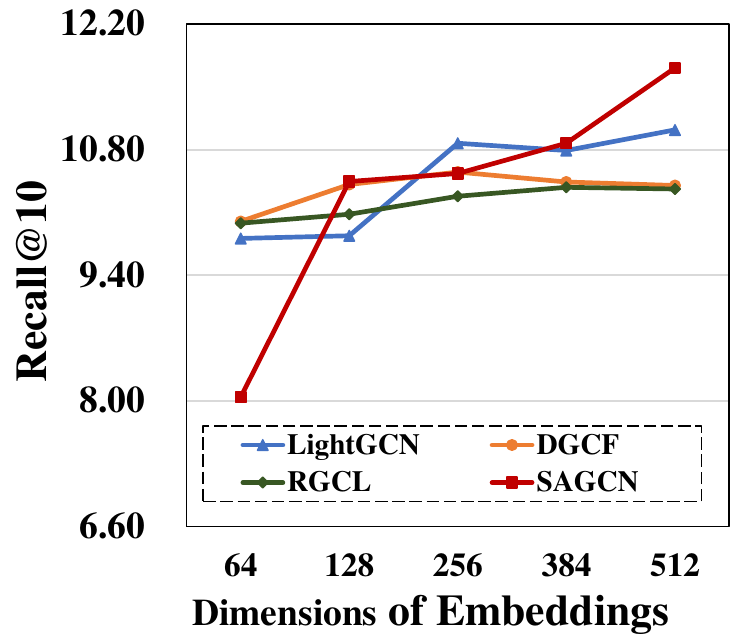}}
         \subfloat[NDCG on Office]{\includegraphics[width=0.25\linewidth]{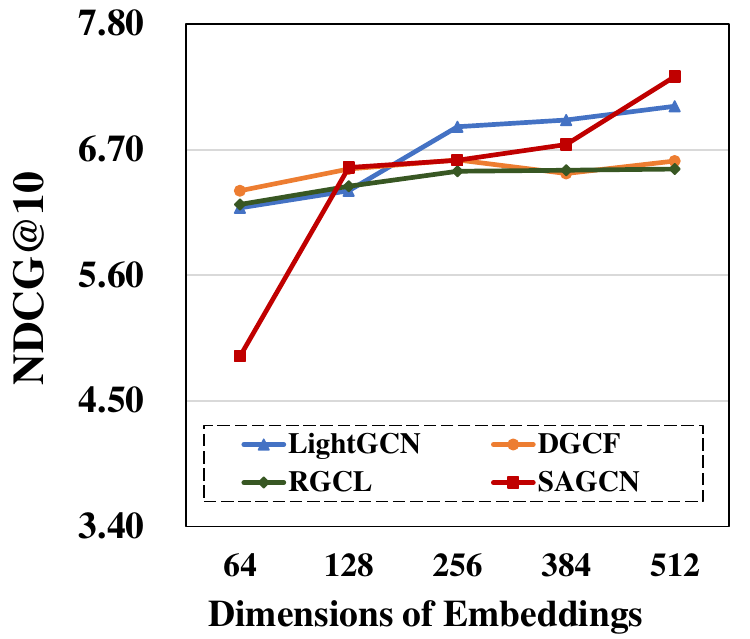}}
          \subfloat[Recall on Clothing]{\includegraphics[width=0.25\linewidth]{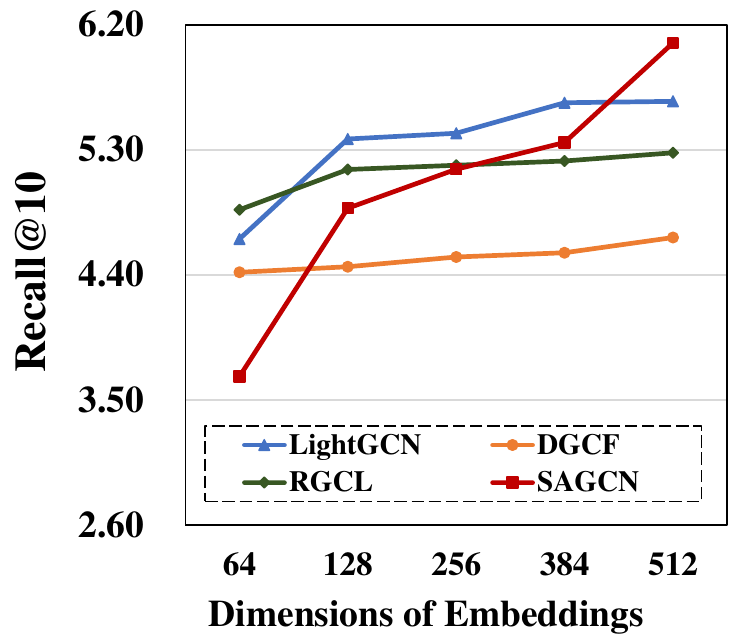}}
         \subfloat[NDCG on Clothing]{\includegraphics[width=0.25\linewidth]{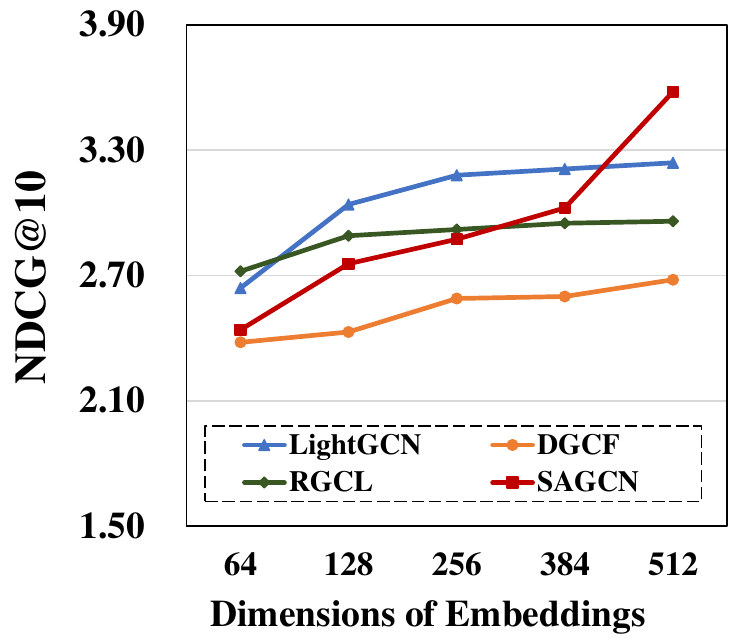}}
         \newline
	\vspace{-0pt}
        \hspace{0mm}
	\caption{{Impact of embedding dimensions on the performance of our proposed method, SAGCN, in comparison with three GCN-based methods (LightGCN, DGCF and RGCL). Notice that the values are reported by percentage with '\%' omitted.}}
\label{fig:EDimmention}
\end{figure*}

{In this section, we analyze the impact of embedding dimensions on the performance of our proposed method, SAGCN, in comparison with three GCN-based methods (LightGCN, DGCF and DCCF). We experiment with embedding dimensions ranging from {64, 128, 256, 384, 512}, evaluating how the size of the embeddings influences model accuracy and overall performance.}

{From the results shown in Fig.~\ref{fig:EDimmention}, we observe a general improvement in performance across all methods as the embedding dimensions increase. However, for LightGCN, DGCF, and RGCL, performance tends to plateau after reaching 128 or 256 dimensions. In contrast, our method, SAGCN, initially performs poorly at 64 dimensions but shows rapid performance growth as the dimensions increase, particularly beyond 384 dimensions.
This improvement can be attributed to the aspect-aware graphs constructed using LLMs to analyze user reviews. Our approach transforms simple user interactions into multiple interaction patterns, significantly enriching the informational content of the interaction data, especially after constructing the aspect-aware graphs. As a result, a larger feature space is required to effectively learn user and item representations from these enriched interactions. The increased information not only enhances the performance of our method but also necessitates a larger feature space to capture the complexity of these interactions.}

\begin{figure*}[t]
	\centering
        \subfloat[Recall on Office]{\includegraphics[width=0.25\linewidth]{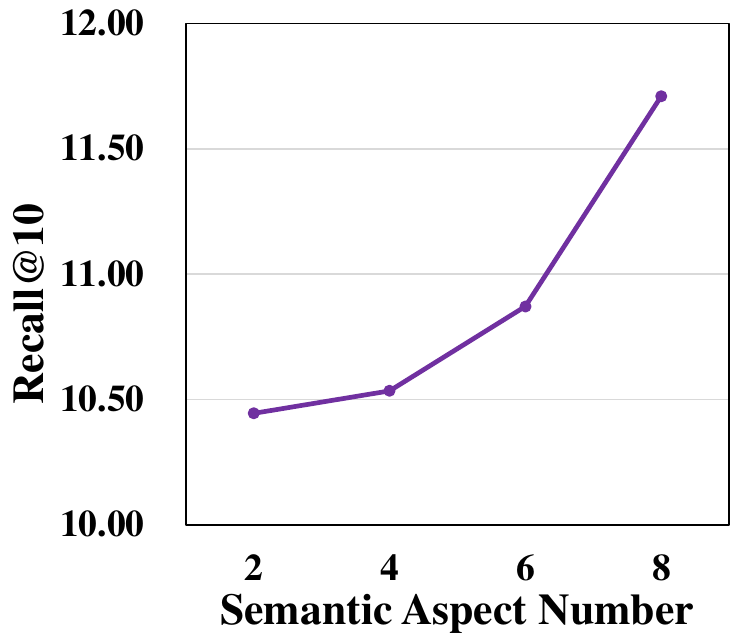}}
        \subfloat[NDCG on Clothing]{\includegraphics[width=0.25\linewidth]{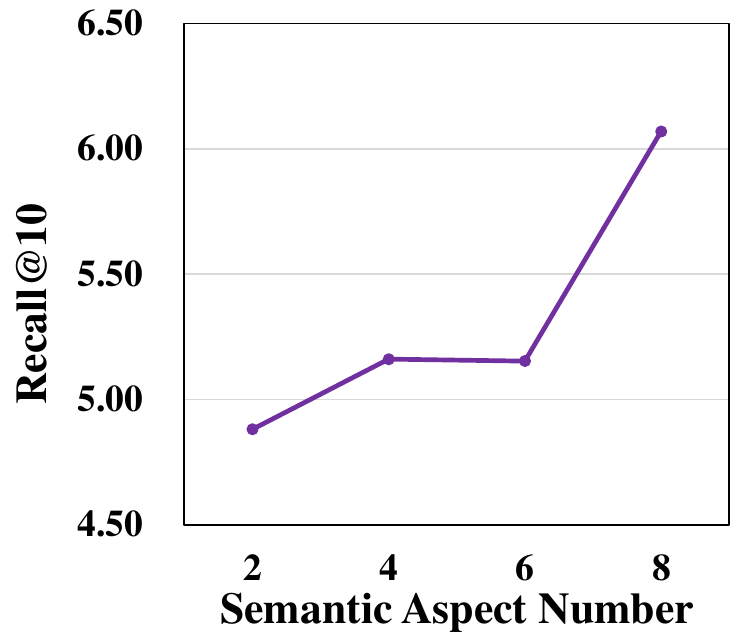}}
        \subfloat[Recall on Baby]{\includegraphics[width=0.25\linewidth]{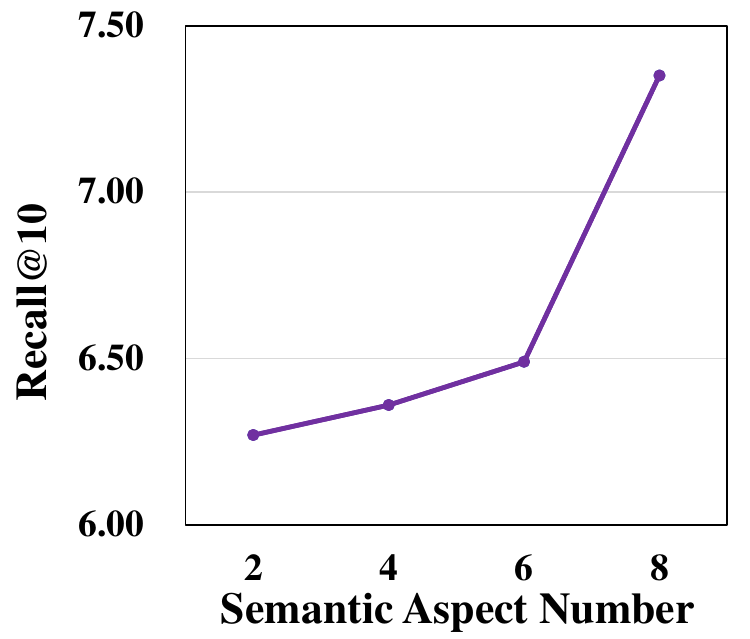}}
        \subfloat[{Recall on Goodreads}]{\includegraphics[width=0.25\linewidth]{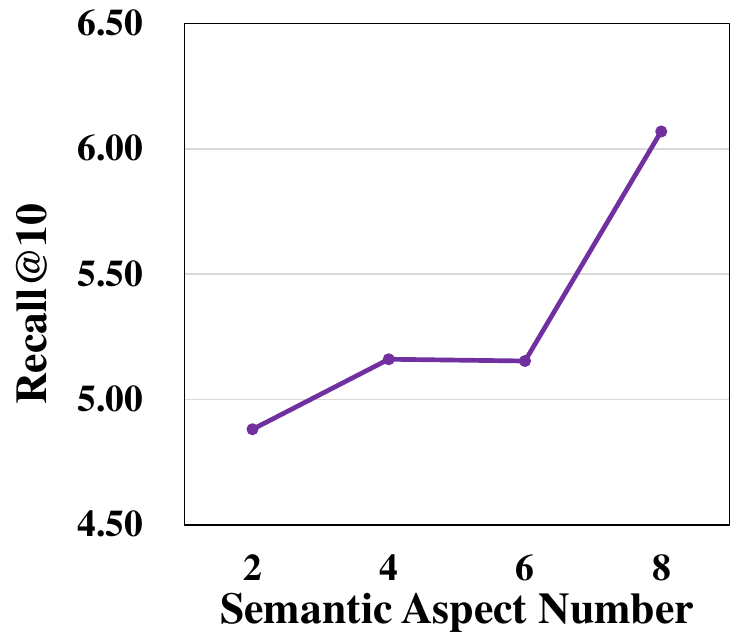}}
        \newline
        \subfloat[NDCG on Office]
        {\includegraphics[width=0.25\linewidth]{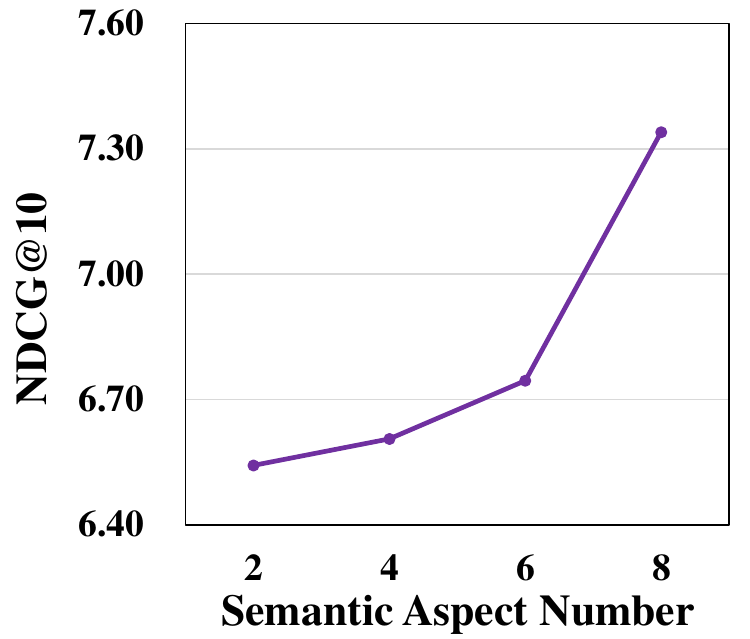}}
        \subfloat[NDCG on Clothing]{\includegraphics[width=0.25\linewidth]{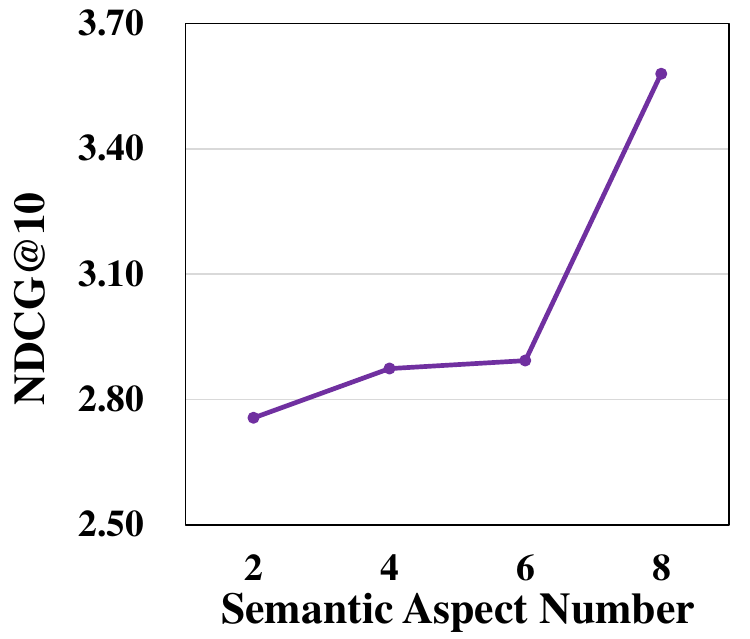}}
        \subfloat[NDCG on Baby]{\includegraphics[width=0.25\linewidth]{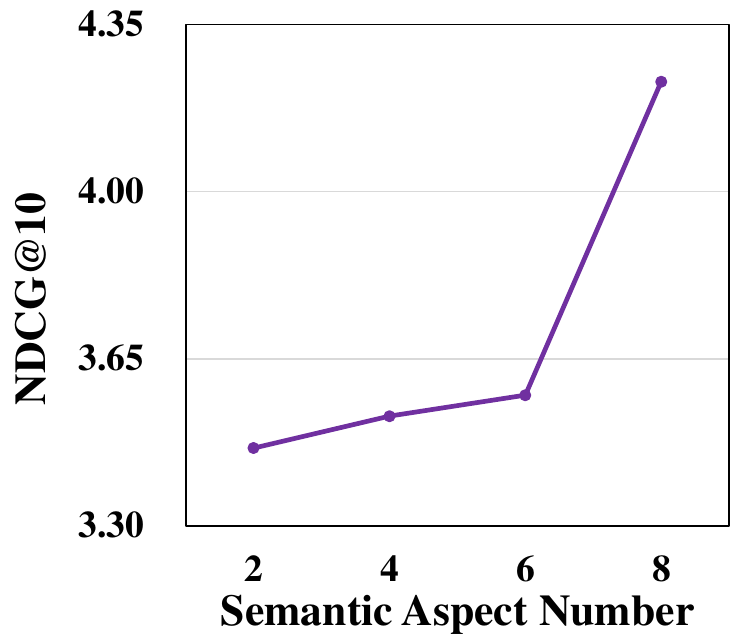}}
        \subfloat[{NDCG on Goodreads}]{\includegraphics[width=0.25\linewidth]{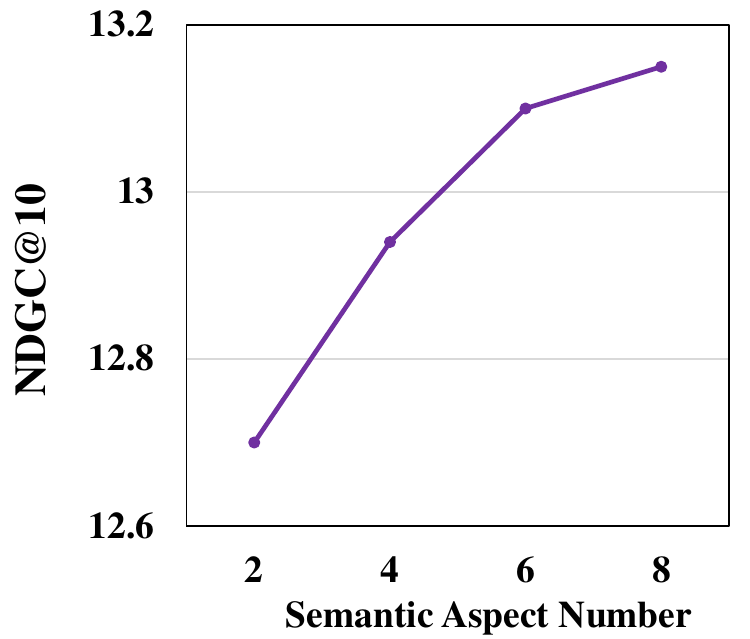}}
        \newline
	\vspace{0pt}
	\caption{{Performance comparison of SAGCN with different semantic aspect numbers on Office and Clothing. Notice that the values are reported by percentage with '\%' omitted.}}
	\vspace{0pt}
\label{fig:SAN}
\end{figure*}

\subsection{\textbf{Semantic Aspects Analysis}}
\begin{table}[t]
\caption{{The extracted semantic-aware aspects and the number of their associated interactions over four datasets.}} 
\resizebox{\textwidth}{18mm}{
\begin{tabular}{c|cccccccc}
\hline
Datasets                  & \multicolumn{8}{c}{Semantic-aware Aspects}                                                                    \\ \hline
\multirow{2}{*}{Office}   & \multicolumn{1}{c|}{\emph{Quality}} & \multicolumn{1}{c|}{\emph{Functionality}} & \multicolumn{1}{c|}{\emph{Ease of Use}} & \multicolumn{1}{c|}{\emph{Convenience}} & \multicolumn{1}{c|}{\emph{Comfort}} & \multicolumn{1}{c|}{\emph{Durability}} & \multicolumn{1}{c|}{\emph{Design}}  & \emph{Price}   \\ \cline{2-9} 
                          & \multicolumn{1}{c|}{43,850}  & \multicolumn{1}{c|}{43,269}        & \multicolumn{1}{c|}{42,347}      & \multicolumn{1}{c|}{41,238}      & \multicolumn{1}{c|}{40,795}  & \multicolumn{1}{c|}{37,564}     & \multicolumn{1}{c|}{23,973}  & 23,661  \\ \hline
\multirow{2}{*}{Baby}     & \multicolumn{1}{c|}{\emph{Quality}} & \multicolumn{1}{c|}{\emph{Functionality}} & \multicolumn{1}{c|}{\emph{Comfort}}     & \multicolumn{1}{c|}{\emph{Ease of Use}} & \multicolumn{1}{c|}{\emph{Design}}  & \multicolumn{1}{c|}{\emph{Durability}} & \multicolumn{1}{c|}{\emph{Size}}    & \emph{Price}   \\ \cline{2-9} 
                          & \multicolumn{1}{c|}{133,337} & \multicolumn{1}{c|}{132,827}       & \multicolumn{1}{c|}{127,938}     & \multicolumn{1}{c|}{125,014}     & \multicolumn{1}{c|}{119,013} & \multicolumn{1}{c|}{116,212}    & \multicolumn{1}{c|}{86,125}  & 59,859  \\ \hline
\multirow{2}{*}{Clothing} & \multicolumn{1}{c|}{\emph{Quality}} & \multicolumn{1}{c|}{\emph{Comfort}}       & \multicolumn{1}{c|}{\emph{Appearance}}  & \multicolumn{1}{c|}{\emph{Style}}       & \multicolumn{1}{c|}{\emph{Fit}}     & \multicolumn{1}{c|}{\emph{Design}}     & \multicolumn{1}{c|}{\emph{Size}}    & \emph{Price}   \\ \cline{2-9} 
                          & \multicolumn{1}{c|}{230,162} & \multicolumn{1}{c|}{210,254}       & \multicolumn{1}{c|}{205,532}     & \multicolumn{1}{c|}{188,378}     & \multicolumn{1}{c|}{186,132} & \multicolumn{1}{c|}{181,222}    & \multicolumn{1}{c|}{170,869} & 106,416 \\ \hline
\multirow{2}{*}{{Goodreads}} & \multicolumn{1}{c|}{\emph{Satisfaction}} & \multicolumn{1}{c|}{\emph{Relevance}}       & \multicolumn{1}{c|}{\emph{Appearance}}  & \multicolumn{1}{c|}{\emph{Support}}       & \multicolumn{1}{c|}{\emph{Price}}     & \multicolumn{1}{c|}{\emph{Usefulness}}     & \multicolumn{1}{c|}{\emph{Quality}}    & \emph{Creativity}   \\ \cline{2-9} 
& \multicolumn{1}{c|}{34,774} & \multicolumn{1}{c|}{31,686}       & \multicolumn{1}{c|}{19,707}     & \multicolumn{1}{c|}{19,366}     & \multicolumn{1}{c|}{14,903} & \multicolumn{1}{c|}{14,111}    & \multicolumn{1}{c|}{11,394} & 10,254 \\ \hline
\end{tabular}}
\label{aspect_statistic}
\vspace{0pt}
\end{table}

\subsubsection{Semantic Aspect-Aware Interactions Analysis}
{In our experiment, we selected eight aspects per dataset by merging similar aspects and eliminating those with very low occurrence.}
Table~\ref{aspect_statistic} reported the number of interactions for different semantic aspects across four datasets. 

Based on our observations of various recommendation scenarios, it is evident that users have distinct concerns regarding different aspects. For instance, in reviews of office products, \emph{convenience} is a common concern, while \emph{fit} is only mentioned in clothing reviews.
To rank the semantic aspects in each recommendation scenario, we analyzed their interaction numbers. This helps us identify that the importance of different semantic aspects varies depending on the number of interactions. For example, in reviews of Office products, \emph{Durability} was mentioned more frequently than \emph{Design}. On the other hand, in reviews of Baby, \emph{Design} was mentioned more frequently than \emph{Durability}. {Note that the most frequently mentioned aspect being \emph{Quality} in the Office, Baby, and Clothing datasets. For these types of products, customers often prioritize \emph{Quality} because it directly influences their decision-making. In contrast, for the GoodReads, the most mentioned aspect is \emph{Satisfaction}, which aligns with the subjective nature of book reviews. Readers tend to express overall satisfaction with the reading experience, which encompasses personal enjoyment, engagement with the story, or alignment with expectations.}

\subsubsection{Effect of Aspect Numbers}
To investigate the effect of semantic aspect numbers, we conducted experiments across all four datasets. Note that all aspects were ranked according to the number of interactions, we selected a number of aspects based on their ranking. Our findings, shown in Fig.~\ref{fig:SAN}, suggest that increasing the semantic aspect numbers can improve the performance of SAGCN. In other words, SAGCN achieves its peak performance when leveraging all aspects, indicating that it can benefit from various aspects. 
Furthermore, we found that SAGCN with eight aspects showed a greater improvement in performance than it with six aspects. This could be due to the fact that the added two aspects might provide more supplementary information to the other six. When combined, the full set of features gives the model a richer understanding of user behaviors.

\subsection{Interpretability Analysis}

\begin{figure*}[t]
    \centering
    \subfloat[Recall on Office]{\includegraphics[width=0.25\linewidth]{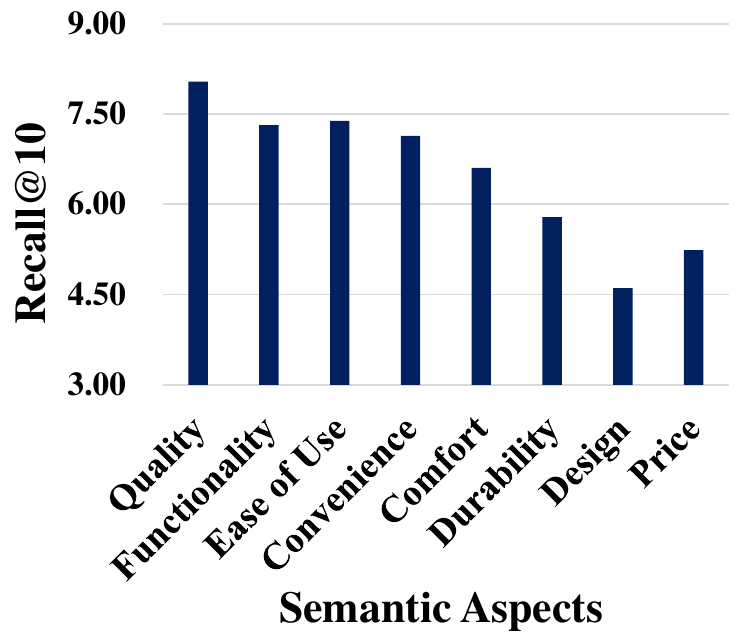}}
    \subfloat[Recall on Clothing]{\includegraphics[width=0.25\linewidth]{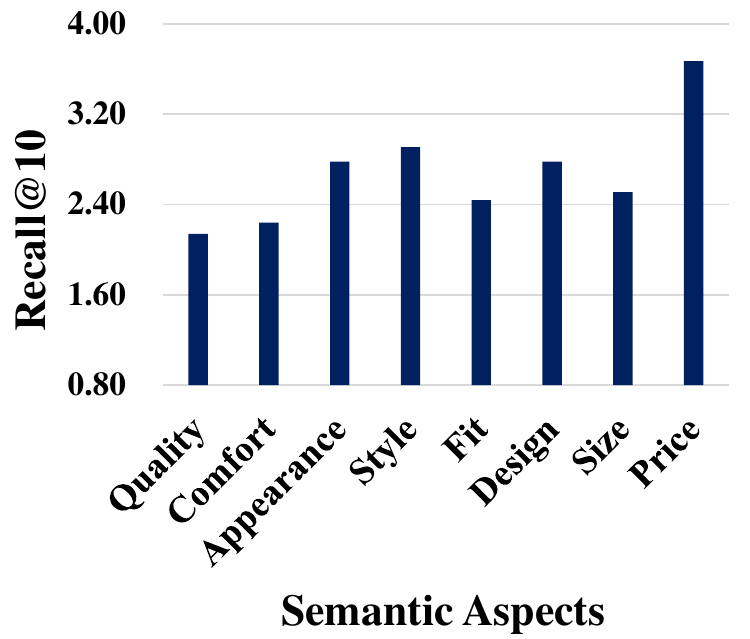}}
    \subfloat[Recall on Baby]{\includegraphics[width=0.25\linewidth]{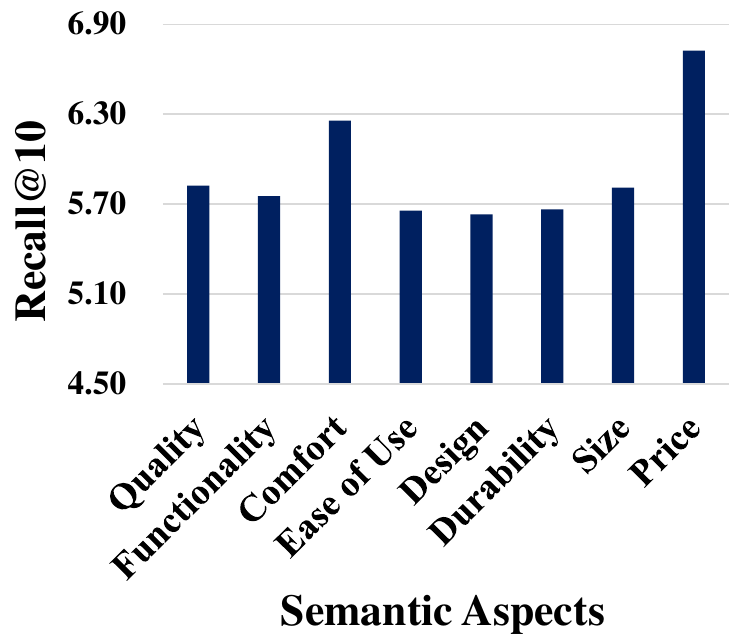}}
    \subfloat[{Recall on Goodreads}]{\includegraphics[width=0.25\linewidth]{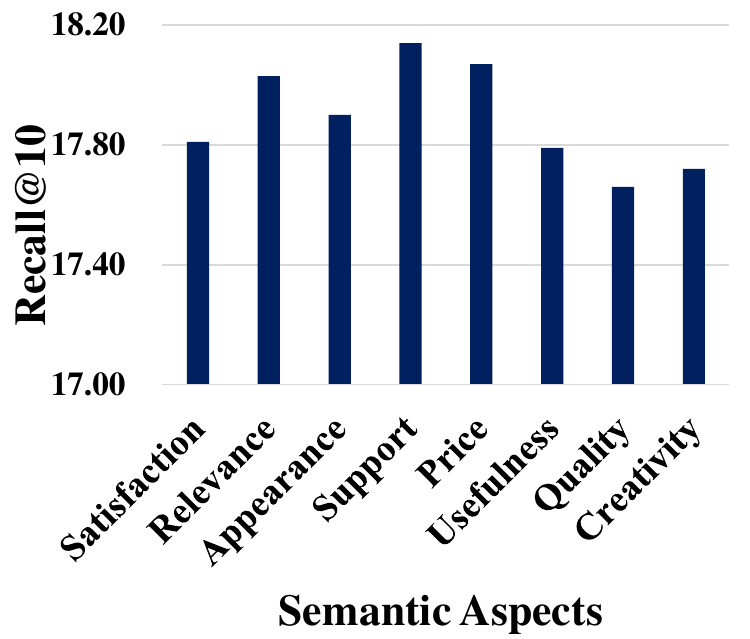}}
    \newline
    \subfloat[NDCG on Office]{\includegraphics[width=0.25\linewidth]{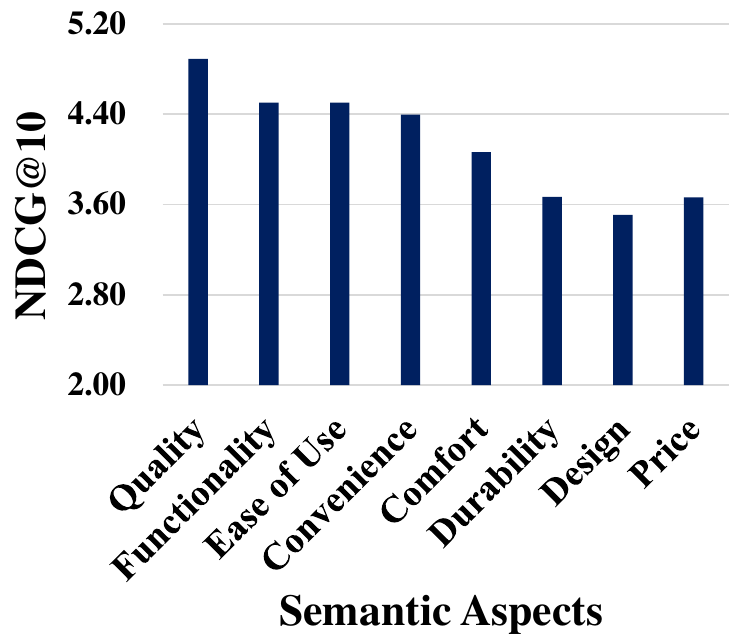}}
    \subfloat[NDCG on Clothing]{\includegraphics[width=0.25\linewidth]{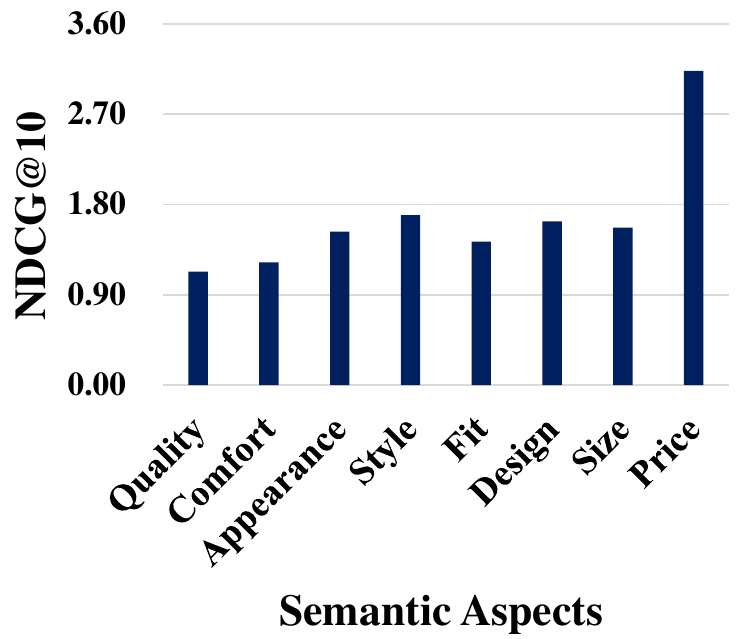}}
    \subfloat[NDCG on Baby]{\includegraphics[width=0.25\linewidth]{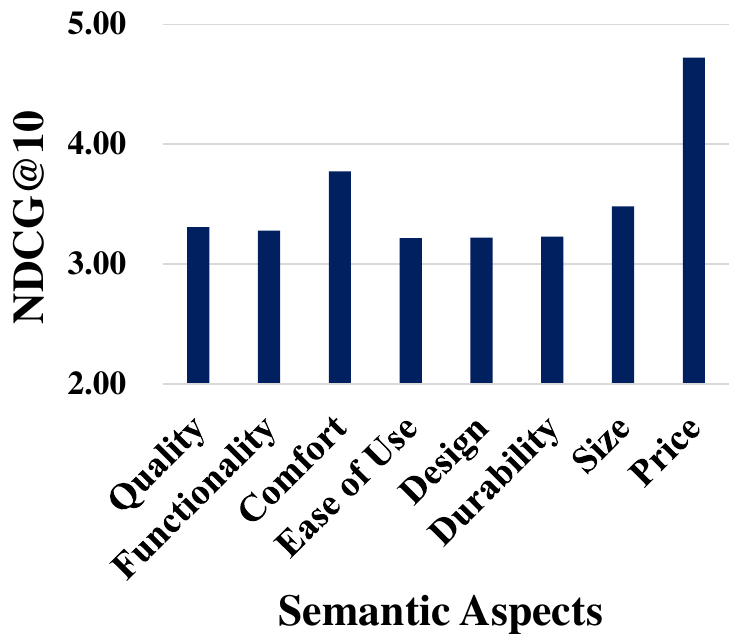}}
    \subfloat[{NDCG on Goodreads}]{\includegraphics[width=0.25\linewidth]{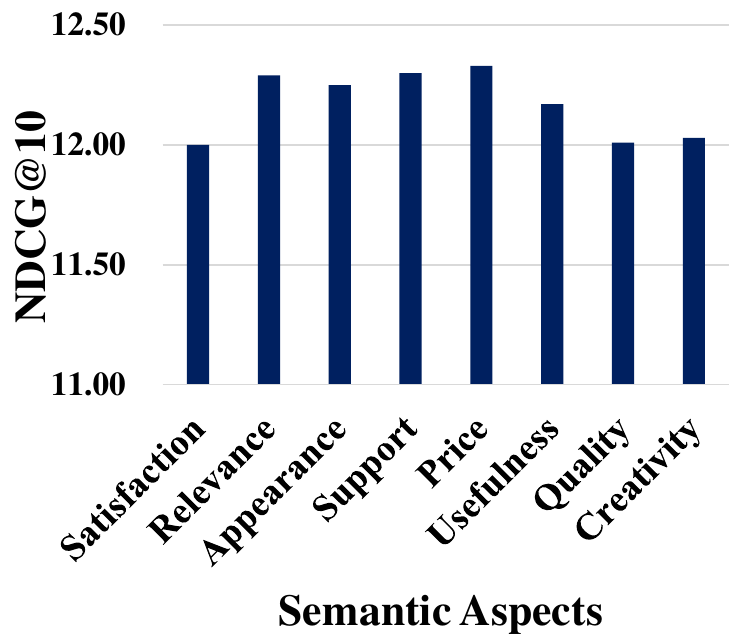}}
    \newline
	\vspace{0pt}
	\caption{{{Performance comparison of SAGCN on different semantic aspects across all four datasets. Notice that the values are reported by percentage with '\%' omitted.}}}
	\vspace{0pt}
\label{fig:SAspects}
\end{figure*}

\begin{figure}[t]
	\centering	
        \subfloat[User (u3547)]{\includegraphics[width=0.5\linewidth]{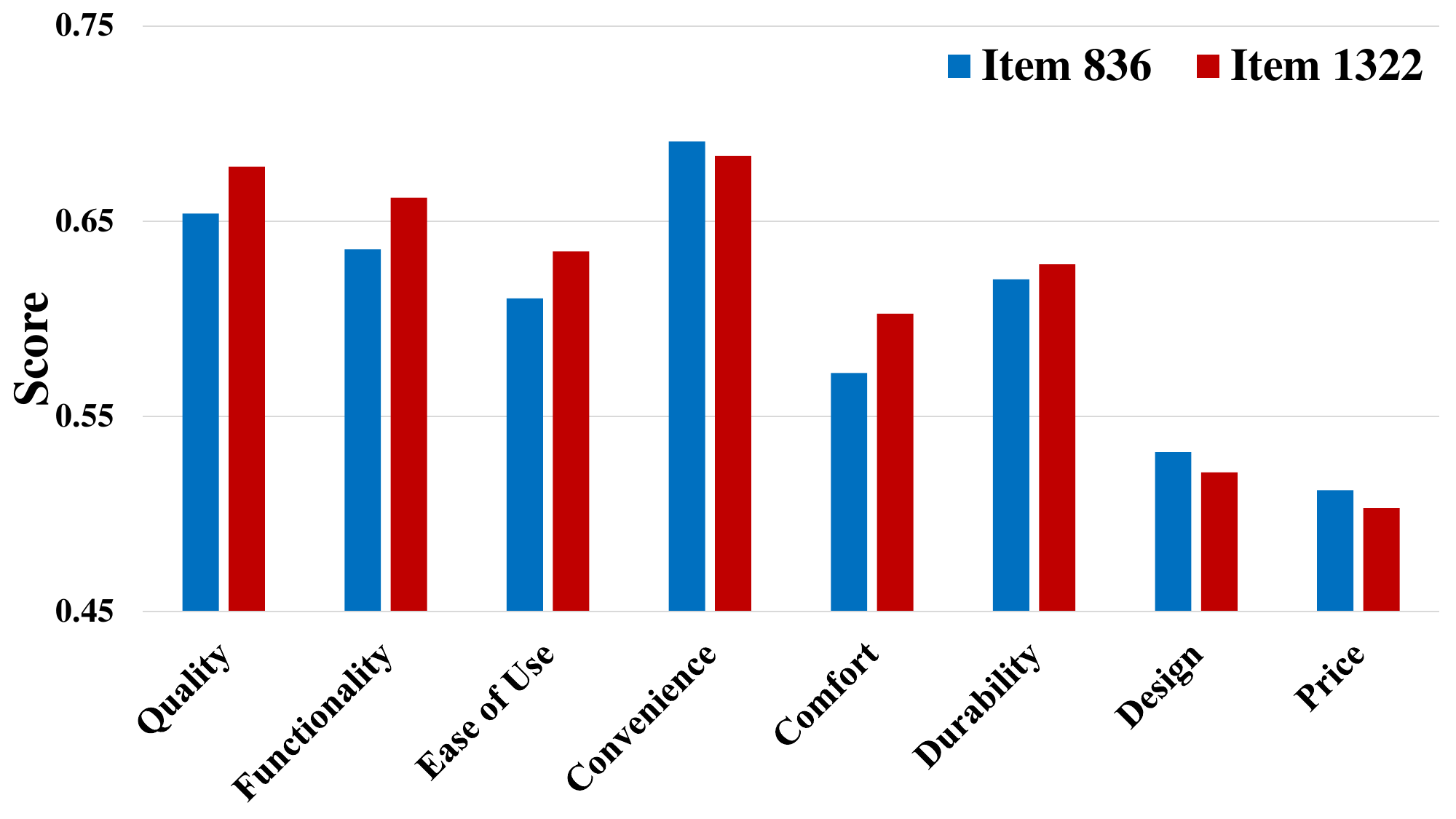}}
        \newline
        \subfloat[Item (i2210)]{\includegraphics[width=0.5\linewidth]{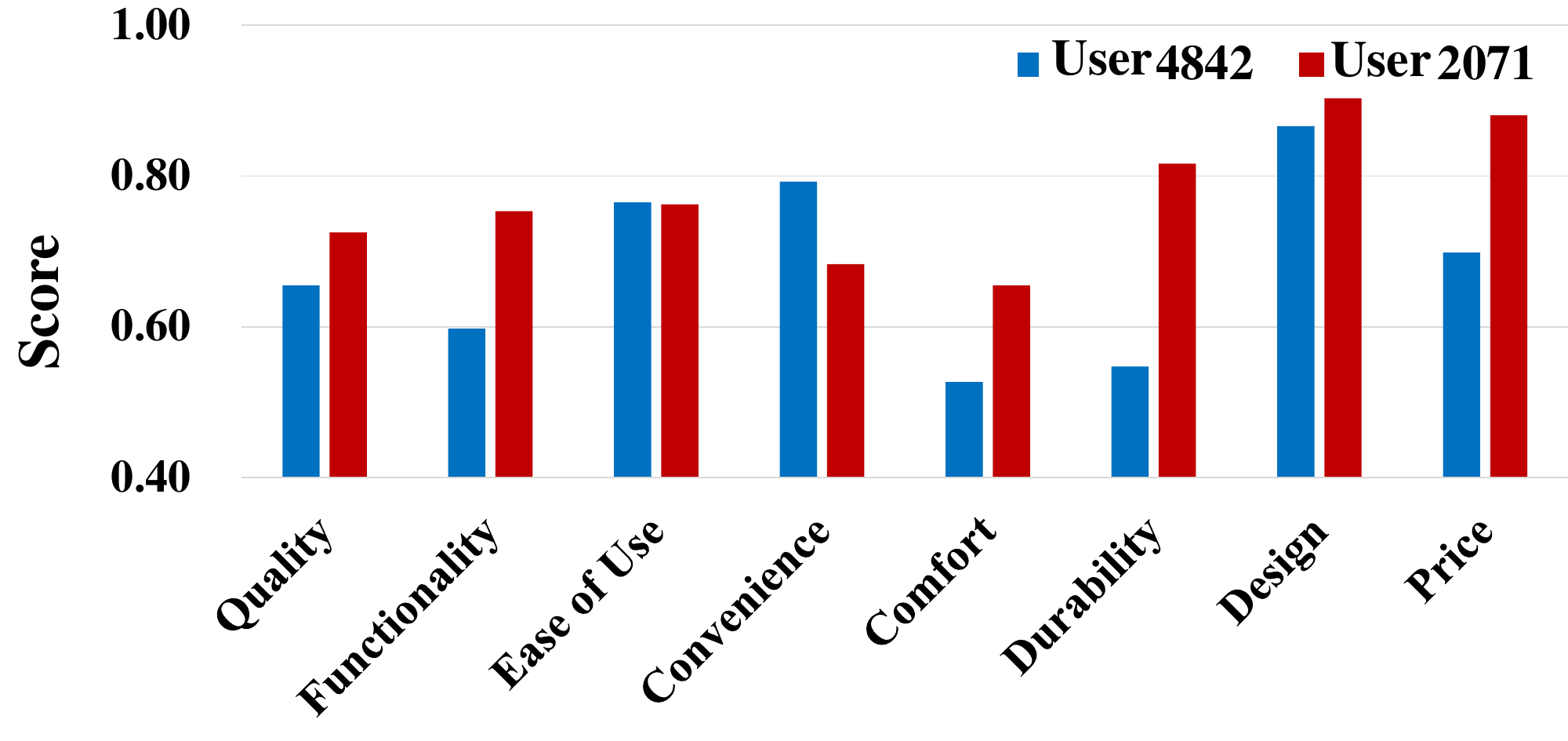}}
        \newline
        \subfloat[User (u3547)]{\includegraphics[width=0.5\linewidth]{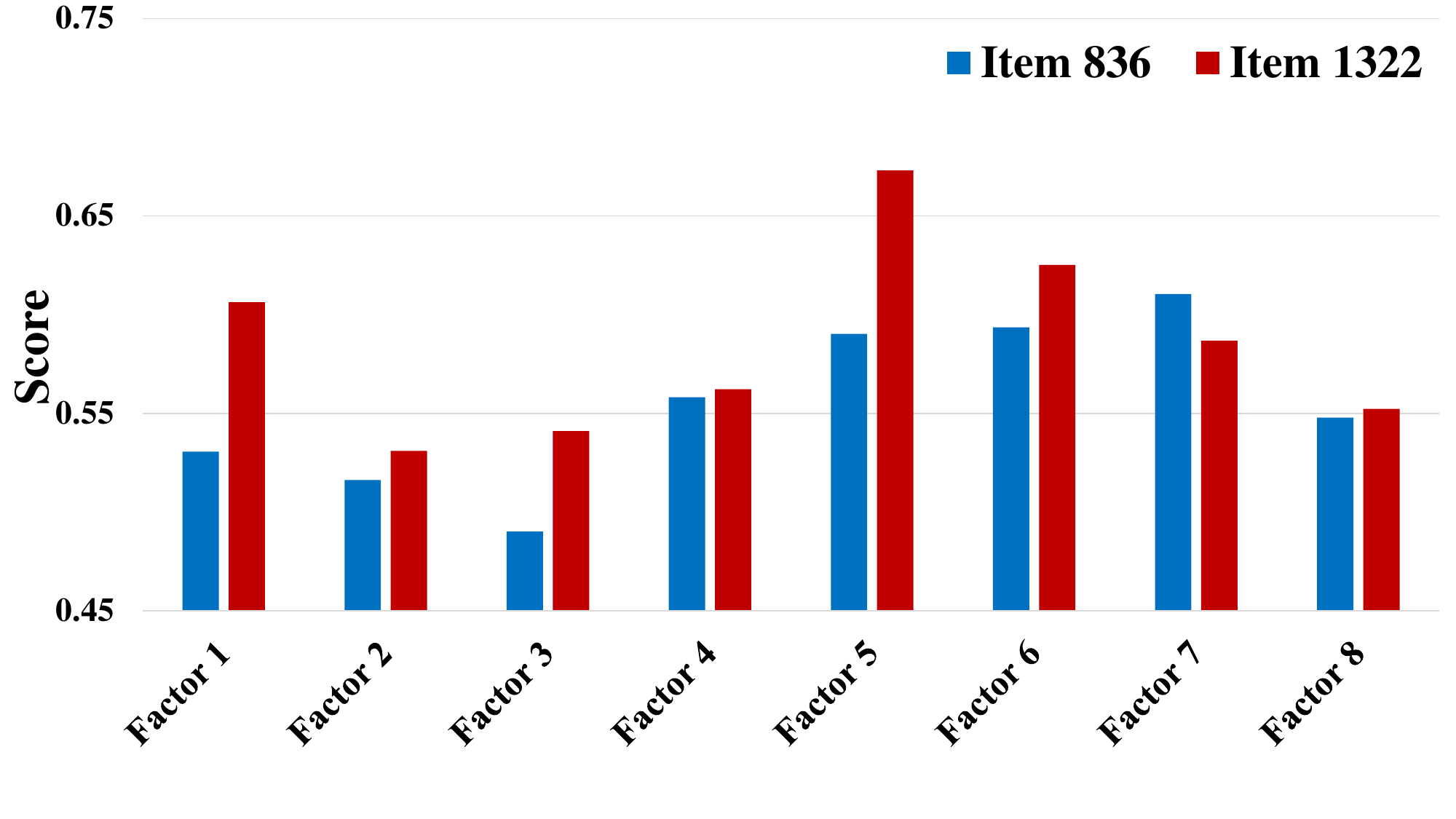}}
        \newline
	\caption{Three examples of preference scores across different aspects in Office. a) Preferences of user $u3547$ for item $i836$ and $i1322$ on different semantic aspects. b) Preferences of user $u4842$ and $u2071$ for item $i2210$ on different semantic aspects. c) Preferences of user $u3547$ for item $i836$ and $i1322$ on different latent factors. Notice that the values are reported by percentage with '\%' omitted.}
	\vspace{0pt}
\label{fig:heatmap}
\end{figure}
In this section, we examined the interpretability of our model from two different perspectives. 
First, our model can interpret the contribution of each semantic aspect in a recommendation scenario. To examine the contribution of different aspects, we predicted the users' preferences using individual aspect embeddings of the users and items. Fig.~\ref{fig:SAspects} shows the following interesting observations: (1) Different semantic aspect features have varying contributions in recommendation. For instance, \emph{Quality} has a greater impact than \emph{Design} for Office. (2) The number of semantic-aware interactions is not the main contributory factor. Surprisingly, even though the number of interactions for \emph{Price} is the least, it still has the most contribution to our model. This could be because \emph{Price} introduces more valuable interactions than the other aspects.

We can further provide a detailed explanation by providing the scores of various semantic aspects for each user-item pair. In Fig.~\ref{fig:heatmap} (a), we computed the preference score of user $u3547$ for items $i836$ and $i1322$ from different semantic aspects in Office. From the result, we can conclude that the user has similar but different preferences on different aspects of different items. Moreover, the user would be interested in these two items mainly due to their \emph{Convenience} and \emph{Quality}. Additionally, the aspect of \emph{Comfort} in $i1322$ might be more appealing to the user than $i836$. Similarly, we computed the preference score of users $u4842$ and $u2071$ for item $i836$, which is illustrated in Figure~\ref{fig:heatmap} (b). These two users have diverse preferences for the same item. Specifically, user $u4842$ focuses on \emph{Design}, \emph{Price} and \emph{Convenience}. In contrast, user $u2071$ pays more attention to the \emph{Design}, \emph{Price}, and \emph{Durability}. These experimental results also demonstrate the interpretability of our proposed approach.

{Overall, we further emphasize our contribution to interpretability compared to traditional methods using disentanglement techniques. As shown in the fig.~\ref{fig:heatmap} (c), in disentanglement-based methods, each representation chunk is associated with a latent factor that cannot be understood by humans. In other words, while we can identify which factor contributes more to user preferences, we have no understanding of what that factor actually represents. In contrast, our approach associates the representations learned from different graphs with specific semantic aspects, which are directly understandable by humans. This demonstrates the advantage of our method in enhancing interpretability, as it provides clear, human-readable explanations for the factors driving user preferences.}

\subsection{Feature Independence Analysis}
\begin{figure*}[t]
	   \centering
         \subfloat[Similarity scores among different aspects (Office)]{\includegraphics[width=0.5\linewidth]{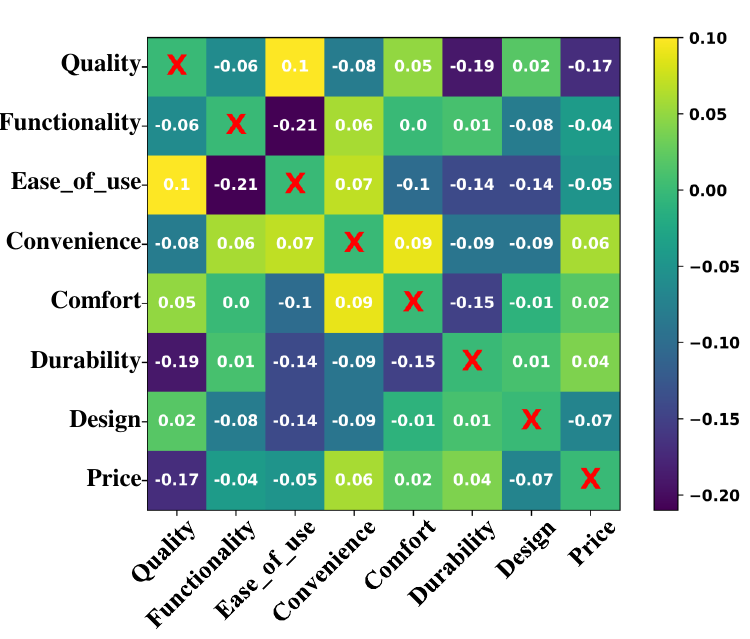}}
         \subfloat[Similarity scores among different aspects (Clothing)]{\includegraphics[width=0.5\linewidth]{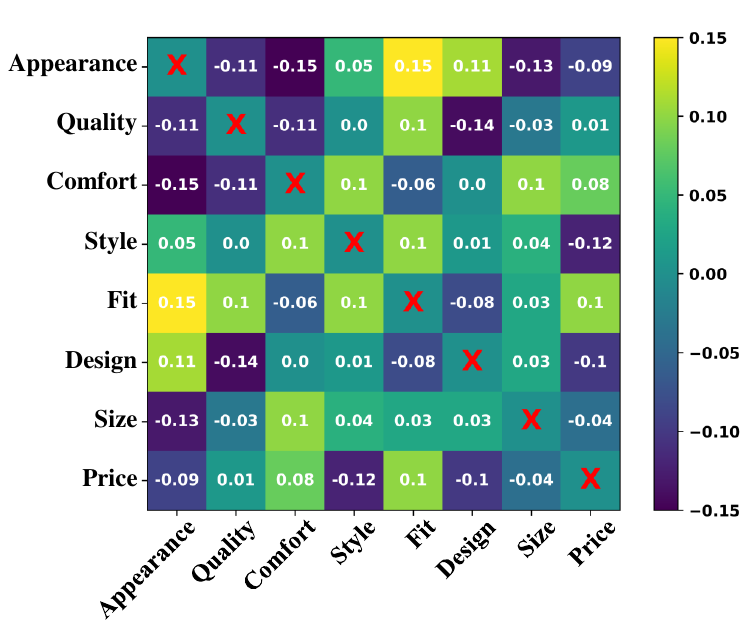}}
         \newline
	\vspace{0pt}
	\caption{Visualization of the feature independence of semantic aspects.}
	\vspace{0pt}
\label{fig:SimAspects}
\end{figure*}

We present a visualization of the feature independence of different semantic aspects in Office and Clothing, as shown in Fig.~\ref{fig:heatmap}. To produce this, we first sampled one user from the user set and then calculated the cosine similarity scores among different aspects based on the user's representation vector. 
The heatmap in Fig.~\ref{fig:heatmap} shows that the similarity values are close to zero between different aspects. This implies that the representations associated with different semantic aspects are independent. This demonstrates that the SAGCN model can learn robust representations with the semantic aspects extracted via LLM, enabling more accurate item recommendations for users. 
An additional noteworthy finding is that the representations are unaffected by the potential bias stemming from the number of semantic aspect-aware interactions. For example, as indicated in Tab.~\ref{aspect_statistic}, there is an imbalance in the number of aspect-aware interactions(e.g., the \emph{Quality} aspect-aware interactions occupy a large proportion in all aspect-aware interactions on both Office and Clothing). In previous data-driven recommendation models, such an imbalance might lead the model to overly focus on these dominant aspects.  However, the SAGCN approach leverages the aspect-aware interactions solely for constructing the semantic-aware graph, meaning that the quantity of these interactions does not sway the aspect-associated representations. This further demonstrates that our proposed model can obtain more robust and better representations by capturing the fine-grain relations between each user and item at the semantic aspect level.

\section{Conclusion and Future Work}
\label{sec:conclusion}
Our paper proposes a novel approach that extracts semantic aspects from user reviews and utilizes them to enhance recommendation accuracy and interpretability. 
Specifically, we introduce a chain-based prompting strategy with LLMs to extract the semantic aspect-aware reviews, which are then utilized to identify semantic aspect-aware user-item interactions. 
To exploit these interactions for representation learning of users and items, we propose a semantic aspect-based graph convolution network (SAGCN) that performs graph convolution operations on multiple semantic aspect-based graphs. 
We conducted extensive experiments on four benchmark datasets to evaluate the performance of our SAGCN model. Our experimental results demonstrate that SAGCN outperforms the pure LLM-based method and the state-of-the-art GCN-based recommendation models. The experiment analysis exhibits the interpretability of our approach.
In summary, our work emphasizes the importance of analyzing user reviews by creating structured reviews to understand users before recommendations. On the other hand, we investigate a new approach to improving recommendation performance by integrating the capability of LLMs with conventional recommendation technique.

This work is an initial attempt to combine the sentiment analysis capabilities of LLMs with conventional recommendation techniques. Our approach involves using a frozen LLM to identify aspect-aware reviews. However, accurately extracting semantic aspects from user reviews remains a challenging task. In addition, apart from the reviews provided by users, the interactions between users and items also include semantic aspects. We expect the potential of sentiment analysis to be further explored for a better understanding of user intent. Furthermore, it would be interesting to explore how to take advantage of these aspect-aware user-item interactions for representation learning.


\bibliographystyle{ACM-Reference-Format}
\bibliography{REF}


\begin{thebibliography}{64}


\ifx \showCODEN    \undefined \def \showCODEN     #1{\unskip}     \fi
\ifx \showDOI      \undefined \def \showDOI       #1{#1}\fi
\ifx \showISBNx    \undefined \def \showISBNx     #1{\unskip}     \fi
\ifx \showISBNxiii \undefined \def \showISBNxiii  #1{\unskip}     \fi
\ifx \showISSN     \undefined \def \showISSN      #1{\unskip}     \fi
\ifx \showLCCN     \undefined \def \showLCCN      #1{\unskip}     \fi
\ifx \shownote     \undefined \def \shownote      #1{#1}          \fi
\ifx \showarticletitle \undefined \def \showarticletitle #1{#1}   \fi
\ifx \showURL      \undefined \def \showURL       {\relax}        \fi
\providecommand\bibfield[2]{#2}
\providecommand\bibinfo[2]{#2}
\providecommand\natexlab[1]{#1}
\providecommand\showeprint[2][]{arXiv:#2}

\bibitem[Bauman et~al\mbox{.}(2017)]%
        {Bauman2017SIGKDD}
\bibfield{author}{\bibinfo{person}{Konstantin Bauman}, \bibinfo{person}{Bing
  Liu}, {and} \bibinfo{person}{Alexander Tuzhilin}.}
  \bibinfo{year}{2017}\natexlab{}.
\newblock \showarticletitle{Aspect Based Recommendations: Recommending Items
  with the Most Valuable Aspects Based on User Reviews}. In
  \bibinfo{booktitle}{\emph{Proceedings of the 23rd ACM SIGKDD International
  Conference on Knowledge Discovery and Data Mining}}.
  \bibinfo{publisher}{ACM}, \bibinfo{pages}{717–725}.
\newblock


\bibitem[Bell and Koren(2007)]%
        {netflix}
\bibfield{author}{\bibinfo{person}{Robert~M. Bell} {and}
  \bibinfo{person}{Yehuda Koren}.} \bibinfo{year}{2007}\natexlab{}.
\newblock \showarticletitle{Lessons from the Netflix prize challenge}.
\newblock \bibinfo{journal}{\emph{ACM SIGKDD Explorations Newsletter}}
  \bibinfo{volume}{9}, \bibinfo{number}{2} (\bibinfo{year}{2007}),
  \bibinfo{pages}{75--79}.
\newblock


\bibitem[Blei et~al\mbox{.}(2003)]%
        {blei2003latent}
\bibfield{author}{\bibinfo{person}{David~M Blei}, \bibinfo{person}{Andrew~Y
  Ng}, {and} \bibinfo{person}{Michael~I Jordan}.}
  \bibinfo{year}{2003}\natexlab{}.
\newblock \showarticletitle{Latent dirichlet allocation}.
\newblock \bibinfo{journal}{\emph{Journal of Machine Learning Research}}
  \bibinfo{volume}{3}, \bibinfo{number}{Jan} (\bibinfo{year}{2003}),
  \bibinfo{pages}{993--1022}.
\newblock


\bibitem[Catherine and Cohen(2017)]%
        {catherine2017transnets}
\bibfield{author}{\bibinfo{person}{Rose Catherine} {and}
  \bibinfo{person}{William Cohen}.} \bibinfo{year}{2017}\natexlab{}.
\newblock \showarticletitle{TransNets: Learning to Transform for
  Recommendation}. In \bibinfo{booktitle}{\emph{Proceedings of the Eleventh ACM
  Conference on Recommender Systems}}. \bibinfo{publisher}{ACM},
  \bibinfo{pages}{288–296}.
\newblock


\bibitem[Chen et~al\mbox{.}(2019)]%
        {chen2019cse}
\bibfield{author}{\bibinfo{person}{Chih{-}Ming Chen},
  \bibinfo{person}{Chuan{-}Ju Wang}, \bibinfo{person}{Ming{-}Feng Tsai}, {and}
  \bibinfo{person}{Yi{-}Hsuan Yang}.} \bibinfo{year}{2019}\natexlab{}.
\newblock \showarticletitle{Collaborative Similarity Embedding for Recommender
  Systems}. In \bibinfo{booktitle}{\emph{The World Wide Web Conference}}.
  \bibinfo{publisher}{ACM}, \bibinfo{pages}{2637--2643}.
\newblock


\bibitem[Chen et~al\mbox{.}(2018)]%
        {chen2018neural}
\bibfield{author}{\bibinfo{person}{Chong Chen}, \bibinfo{person}{Min Zhang},
  \bibinfo{person}{Yiqun Liu}, {and} \bibinfo{person}{Shaoping Ma}.}
  \bibinfo{year}{2018}\natexlab{}.
\newblock \showarticletitle{Neural attentional rating regression with
  review-level explanations}. In \bibinfo{booktitle}{\emph{Proceedings of the
  2018 World Wide Web Conference}}. IW3C2, \bibinfo{pages}{1583--1592}.
\newblock


\bibitem[Chen et~al\mbox{.}(2024)]%
        {Chen2024CDR}
\bibfield{author}{\bibinfo{person}{Hong Chen}, \bibinfo{person}{Yudong Chen},
  \bibinfo{person}{Xin Wang}, \bibinfo{person}{Ruobing Xie},
  \bibinfo{person}{Rui Wang}, \bibinfo{person}{Feng Xia}, {and}
  \bibinfo{person}{Wenwu Zhu}.} \bibinfo{year}{2024}\natexlab{}.
\newblock \showarticletitle{Curriculum disentangled recommendation with noisy
  multi-feedback}. In \bibinfo{booktitle}{\emph{Proceedings of the 35th
  International Conference on Neural Information Processing Systems}}.
  \bibinfo{publisher}{CAI.}, \bibinfo{pages}{26924--26936}.
\newblock


\bibitem[Cheng et~al\mbox{.}(2018a)]%
        {cheng20183ncf}
\bibfield{author}{\bibinfo{person}{Zhiyong Cheng}, \bibinfo{person}{Ying Ding},
  \bibinfo{person}{Xiangnan He}, \bibinfo{person}{Lei Zhu},
  \bibinfo{person}{Xuemeng Song}, {and} \bibinfo{person}{Mohan Kankanhalli}.}
  \bibinfo{year}{2018}\natexlab{a}.
\newblock \showarticletitle{A$^3$NCF: An Adaptive Aspect Attention Model for
  Rating Prediction}. In \bibinfo{booktitle}{\emph{Proceedings of the 27th
  International Joint Conference on Artificial Intelligence}}.
  \bibinfo{publisher}{AAAI Press}, \bibinfo{pages}{3748--3754}.
\newblock


\bibitem[Cheng et~al\mbox{.}(2018b)]%
        {cheng2018aspect}
\bibfield{author}{\bibinfo{person}{Zhiyong Cheng}, \bibinfo{person}{Ying Ding},
  \bibinfo{person}{Lei Zhu}, {and} \bibinfo{person}{Mohan Kankanhalli}.}
  \bibinfo{year}{2018}\natexlab{b}.
\newblock \showarticletitle{Aspect-aware latent factor model: Rating prediction
  with ratings and reviews}. In \bibinfo{booktitle}{\emph{Proceedings of The
  Web Conference 2018}}. \bibinfo{publisher}{IW3C2}, \bibinfo{pages}{639--648}.
\newblock


\bibitem[Cheng et~al\mbox{.}(2023)]%
        {Cheng2023MBR}
\bibfield{author}{\bibinfo{person}{Zhiyong Cheng}, \bibinfo{person}{Sai Han},
  \bibinfo{person}{Fan Liu}, \bibinfo{person}{Lei Zhu}, \bibinfo{person}{Zan
  Gao}, {and} \bibinfo{person}{Yuxin Peng}.} \bibinfo{year}{2023}\natexlab{}.
\newblock \showarticletitle{Multi-Behavior Recommendation with Cascading Graph
  Convolution Networks}. In \bibinfo{booktitle}{\emph{Proceedings of the ACM
  Web Conference 2023}}. \bibinfo{publisher}{ACM},
  \bibinfo{pages}{1181–1189}.
\newblock


\bibitem[Chin et~al\mbox{.}(2018)]%
        {chin2018anr}
\bibfield{author}{\bibinfo{person}{Jin~Yao Chin}, \bibinfo{person}{Kaiqi Zhao},
  \bibinfo{person}{Shafiq Joty}, {and} \bibinfo{person}{Gao Cong}.}
  \bibinfo{year}{2018}\natexlab{}.
\newblock \showarticletitle{{ANR}: Aspect-based Neural Recommender}. In
  \bibinfo{booktitle}{\emph{Proceedings of the 2018 ACM on Conference on
  Information and Knowledge Management}}. ACM, \bibinfo{pages}{147--156}.
\newblock


\bibitem[Da’u and Salim(2022)]%
        {Da2020AIR}
\bibfield{author}{\bibinfo{person}{Aminu Da’u} {and} \bibinfo{person}{Naomie
  Salim}.} \bibinfo{year}{2022}\natexlab{}.
\newblock \showarticletitle{Recommendation system based on deep learning
  methods: a systematic review and new directions}.
\newblock \bibinfo{journal}{\emph{Artificial Intelligence Review}}
  \bibinfo{volume}{53}, \bibinfo{number}{4} (\bibinfo{year}{2022}),
  \bibinfo{pages}{2709--2748}.
\newblock


\bibitem[Devlin et~al\mbox{.}(2019)]%
        {Devlin2019BERTPO}
\bibfield{author}{\bibinfo{person}{Jacob Devlin}, \bibinfo{person}{Ming-Wei
  Chang}, \bibinfo{person}{Kenton Lee}, {and} \bibinfo{person}{Kristina
  Toutanova}.} \bibinfo{year}{2019}\natexlab{}.
\newblock \showarticletitle{BERT: Pre-training of Deep Bidirectional
  Transformers for Language Understanding}. In
  \bibinfo{booktitle}{\emph{Proceedings of the 2019 Conference of the North
  American Chapter of the Association for Computational Linguistics: Human
  Language Technologies}}. \bibinfo{publisher}{ACL},
  \bibinfo{pages}{4171--4186}.
\newblock


\bibitem[Fan et~al\mbox{.}(2022)]%
        {fan2022GTN}
\bibfield{author}{\bibinfo{person}{Wenqi Fan}, \bibinfo{person}{Xiaorui Liu},
  \bibinfo{person}{Wei Jin}, \bibinfo{person}{Xiangyu Zhao},
  \bibinfo{person}{Jiliang Tang}, {and} \bibinfo{person}{Qing Li}.}
  \bibinfo{year}{2022}\natexlab{}.
\newblock \showarticletitle{Graph Trend Filtering Networks for Recommendation}.
  In \bibinfo{booktitle}{\emph{Proceedings of the 45th International ACM SIGIR
  Conference on Research and Development in Information Retrieval}}.
  \bibinfo{pages}{112--121}.
\newblock


\bibitem[Fouss et~al\mbox{.}(2007)]%
        {fouss2007tkde}
\bibfield{author}{\bibinfo{person}{Fran{\c{c}}ois Fouss},
  \bibinfo{person}{Alain Pirotte}, \bibinfo{person}{Jean{-}Michel Renders},
  {and} \bibinfo{person}{Marco Saerens}.} \bibinfo{year}{2007}\natexlab{}.
\newblock \showarticletitle{Random-Walk Computation of Similarities between
  Nodes of a Graph with Application to Collaborative Recommendation}.
\newblock \bibinfo{journal}{\emph{Transactions on Knowledge and Data
  Engineering}} \bibinfo{volume}{19}, \bibinfo{number}{3}
  (\bibinfo{year}{2007}), \bibinfo{pages}{355--369}.
\newblock


\bibitem[Geng et~al\mbox{.}(2022)]%
        {RLP2022RecSys}
\bibfield{author}{\bibinfo{person}{Shijie Geng}, \bibinfo{person}{Shuchang
  Liu}, \bibinfo{person}{Zuohui Fu}, \bibinfo{person}{Yingqiang Ge}, {and}
  \bibinfo{person}{Yongfeng Zhang}.} \bibinfo{year}{2022}\natexlab{}.
\newblock \showarticletitle{Recommendation as language processing (rlp): A
  unified pretrain, personalized prompt \& predict paradigm (p5)}. In
  \bibinfo{booktitle}{\emph{Proceedings of the 16th ACM Conference on
  Recommender Systems}}. \bibinfo{pages}{299--315}.
\newblock


\bibitem[Gori and Pucci(2007)]%
        {gori2007itemrank}
\bibfield{author}{\bibinfo{person}{Marco Gori} {and} \bibinfo{person}{Augusto
  Pucci}.} \bibinfo{year}{2007}\natexlab{}.
\newblock \showarticletitle{ItemRank: A Random-Walk Based Scoring Algorithm for
  Recommender Engines}. In \bibinfo{booktitle}{\emph{Proceedings of the 20th
  International Joint Conference on Artifical Intelligence}}.
  \bibinfo{publisher}{MKPI}, \bibinfo{pages}{2766–2771}.
\newblock


\bibitem[He et~al\mbox{.}(2015)]%
        {he2015trirank}
\bibfield{author}{\bibinfo{person}{Xiangnan He}, \bibinfo{person}{Tao Chen},
  \bibinfo{person}{Min-Yen Kan}, {and} \bibinfo{person}{Xiao Chen}.}
  \bibinfo{year}{2015}\natexlab{}.
\newblock \showarticletitle{TriRank: Review aware Explainable Recommendation by
  Modeling Aspects}. In \bibinfo{booktitle}{\emph{Proceedings of the 2015 ACM
  on Conference on Information and Knowledge Management}}.
  \bibinfo{publisher}{ACM}, \bibinfo{pages}{1661--1670}.
\newblock


\bibitem[He et~al\mbox{.}(2020)]%
        {He@LightGCN}
\bibfield{author}{\bibinfo{person}{Xiangnan He}, \bibinfo{person}{Kuan Deng},
  \bibinfo{person}{Xiang Wang}, \bibinfo{person}{Yan Li},
  \bibinfo{person}{YongDong Zhang}, {and} \bibinfo{person}{Meng Wang}.}
  \bibinfo{year}{2020}\natexlab{}.
\newblock \showarticletitle{LightGCN: Simplifying and Powering Graph
  Convolution Network for Recommendation}. In
  \bibinfo{booktitle}{\emph{Proceedings of the 43rd International ACM SIGIR
  Conference on Research and Development in Information Retrieval}}.
  \bibinfo{publisher}{ACM}, \bibinfo{pages}{639–648}.
\newblock


\bibitem[He et~al\mbox{.}(2017)]%
        {he2017neural}
\bibfield{author}{\bibinfo{person}{Xiangnan He}, \bibinfo{person}{Lizi Liao},
  \bibinfo{person}{Hanwang Zhang}, \bibinfo{person}{Liqiang Nie},
  \bibinfo{person}{Xia Hu}, {and} \bibinfo{person}{Tat-Seng Chua}.}
  \bibinfo{year}{2017}\natexlab{}.
\newblock \showarticletitle{Neural collaborative filtering}. In
  \bibinfo{booktitle}{\emph{Proceedings of the 26th International Conference on
  World Wide Web}}. \bibinfo{publisher}{IW3C2}, \bibinfo{pages}{173--182}.
\newblock


\bibitem[Ji et~al\mbox{.}(2024)]%
        {Ji2023GenRec}
\bibfield{author}{\bibinfo{person}{Jianchao Ji}, \bibinfo{person}{Zelong Li},
  \bibinfo{person}{Shuyuan Xu}, \bibinfo{person}{Wenyue Hua},
  \bibinfo{person}{Yingqiang Ge}, \bibinfo{person}{Juntao Tan}, {and}
  \bibinfo{person}{Yongfeng Zhang}.} \bibinfo{year}{2024}\natexlab{}.
\newblock \showarticletitle{GenRec: Large Language Model for Generative
  Recommendation}. In \bibinfo{booktitle}{\emph{Proceedings of the 2024 Joint
  International Conference on Computational Linguistics, Language Resources and
  Evaluation}}. \bibinfo{publisher}{ELRA and ICCL},
  \bibinfo{pages}{10146--10159}.
\newblock


\bibitem[Kang et~al\mbox{.}(2023)]%
        {wang2023LLM}
\bibfield{author}{\bibinfo{person}{Wang{-}Cheng Kang}, \bibinfo{person}{Jianmo
  Ni}, \bibinfo{person}{Nikhil Mehta}, \bibinfo{person}{Maheswaran
  Sathiamoorthy}, \bibinfo{person}{Lichan Hong}, \bibinfo{person}{Ed~H. Chi},
  {and} \bibinfo{person}{Derek~Zhiyuan Cheng}.}
  \bibinfo{year}{2023}\natexlab{}.
\newblock \showarticletitle{Do LLMs Understand User Preferences? Evaluating
  LLMs On User Rating Prediction}. In \bibinfo{booktitle}{\emph{Proceedings of
  the Personalized Generative AI Workshop 2023}}. \bibinfo{publisher}{ACM},
  \bibinfo{pages}{1--11}.
\newblock


\bibitem[Kingma and Ba(2015)]%
        {kingma2014adam}
\bibfield{author}{\bibinfo{person}{Diederik~P Kingma} {and}
  \bibinfo{person}{Jimmy Ba}.} \bibinfo{year}{2015}\natexlab{}.
\newblock \showarticletitle{Adam: A method for stochastic optimization}. In
  \bibinfo{booktitle}{\emph{Proceedings of the 3rd International Conference on
  Learning Representations}}. \bibinfo{publisher}{OpenReview.net},
  \bibinfo{pages}{1--15}.
\newblock


\bibitem[Kipf and Welling(2017)]%
        {kipf2017gcn}
\bibfield{author}{\bibinfo{person}{Thomas~N. Kipf} {and} \bibinfo{person}{Max
  Welling}.} \bibinfo{year}{2017}\natexlab{}.
\newblock \showarticletitle{Semi-Supervised Classification with Graph
  Convolutional Networks}. In \bibinfo{booktitle}{\emph{Proceedings of the 5th
  International Conference on Learning Representations}}.
  \bibinfo{publisher}{OpenReview.net}, \bibinfo{pages}{1--14}.
\newblock


\bibitem[Koren et~al\mbox{.}(2009)]%
        {Koren2009MF}
\bibfield{author}{\bibinfo{person}{Yehuda Koren}, \bibinfo{person}{Robert
  Bell}, {and} \bibinfo{person}{Chris Volinsky}.}
  \bibinfo{year}{2009}\natexlab{}.
\newblock \showarticletitle{Matrix factorization techniques for recommender
  systems}.
\newblock \bibinfo{journal}{\emph{Computer}} \bibinfo{volume}{42},
  \bibinfo{number}{8} (\bibinfo{year}{2009}), \bibinfo{pages}{30--37}.
\newblock


\bibitem[Li et~al\mbox{.}(2019)]%
        {Li2019CNR}
\bibfield{author}{\bibinfo{person}{Chenliang Li}, \bibinfo{person}{Cong Quan},
  \bibinfo{person}{Li Peng}, \bibinfo{person}{Yunwei Qi},
  \bibinfo{person}{Yuming Deng}, {and} \bibinfo{person}{Libing Wu}.}
  \bibinfo{year}{2019}\natexlab{}.
\newblock \showarticletitle{A Capsule Network for Recommendation and Explaining
  What You Like and Dislike}. In \bibinfo{booktitle}{\emph{Proceedings of the
  42nd International ACM SIGIR Conference on Research and Development in
  Information Retrieval}}. \bibinfo{publisher}{ACM},
  \bibinfo{pages}{275–284}.
\newblock


\bibitem[Li et~al\mbox{.}(2023)]%
        {Li2023PEPLER}
\bibfield{author}{\bibinfo{person}{Lei Li}, \bibinfo{person}{Yongfeng Zhang},
  {and} \bibinfo{person}{Li Chen}.} \bibinfo{year}{2023}\natexlab{}.
\newblock \showarticletitle{Personalized Prompt Learning for Explainable
  Recommendation}.
\newblock \bibinfo{journal}{\emph{ACM Transactions on Information Systems}}
  \bibinfo{volume}{41}, \bibinfo{number}{4} (\bibinfo{year}{2023}),
  \bibinfo{pages}{1--26}.
\newblock


\bibitem[Li et~al\mbox{.}(2024)]%
        {ADDRL@MM}
\bibfield{author}{\bibinfo{person}{Zhenyang Li}, \bibinfo{person}{Fan Liu},
  \bibinfo{person}{Yinwei Wei}, \bibinfo{person}{Zhiyong Cheng},
  \bibinfo{person}{Liqiang Nie}, {and} \bibinfo{person}{Mohan Kankanhalli}.}
  \bibinfo{year}{2024}\natexlab{}.
\newblock \showarticletitle{Attribute-driven Disentangled Representation
  Learning for Multimodal Recommendation}. In
  \bibinfo{booktitle}{\emph{Proceedings of the 32nd ACM International
  Conference on Multimedia}}. \bibinfo{publisher}{ACM},
  \bibinfo{pages}{9660–9669}.
\newblock


\bibitem[Lin et~al\mbox{.}(2022)]%
        {lin2022NCL}
\bibfield{author}{\bibinfo{person}{Zihan Lin}, \bibinfo{person}{Changxin Tian},
  \bibinfo{person}{Yupeng Hou}, {and} \bibinfo{person}{Wayne~Xin Zhao}.}
  \bibinfo{year}{2022}\natexlab{}.
\newblock \showarticletitle{Improving graph collaborative filtering with
  neighborhood-enriched contrastive learning}. In
  \bibinfo{booktitle}{\emph{Proceedings of the ACM Web Conference 2022}}.
  \bibinfo{publisher}{ACM}, \bibinfo{pages}{2320--2329}.
\newblock


\bibitem[Liu et~al\mbox{.}(2023a)]%
        {Liu2022DMRL}
\bibfield{author}{\bibinfo{person}{Fan Liu}, \bibinfo{person}{Huilin Chen},
  \bibinfo{person}{Zhiyong Cheng}, \bibinfo{person}{Anan Liu},
  \bibinfo{person}{Liqiang Nie}, {and} \bibinfo{person}{Mohan Kankanhalli}.}
  \bibinfo{year}{2023}\natexlab{a}.
\newblock \showarticletitle{Disentangled Multimodal Representation Learning for
  Recommendation}.
\newblock \bibinfo{journal}{\emph{IEEE Transactions on Multimedia}}
  \bibinfo{volume}{25} (\bibinfo{year}{2023}), \bibinfo{pages}{7149--7159}.
\newblock


\bibitem[Liu et~al\mbox{.}(2019)]%
        {liu2018MAML}
\bibfield{author}{\bibinfo{person}{Fan Liu}, \bibinfo{person}{Zhiyong Cheng},
  \bibinfo{person}{Changchang Sun}, \bibinfo{person}{Yinglong Wang},
  \bibinfo{person}{Liqiang Nie}, {and} \bibinfo{person}{Mohan Kankanhalli}.}
  \bibinfo{year}{2019}\natexlab{}.
\newblock \showarticletitle{User Diverse Preference Modeling by Multimodal
  Attentive Metric Learning}. In \bibinfo{booktitle}{\emph{Proceedings of the
  27th ACM International Conference on Multimedia}}. \bibinfo{publisher}{ACM},
  \bibinfo{pages}{1526–1534}.
\newblock


\bibitem[Liu et~al\mbox{.}(2021)]%
        {Liu2021IMP_GCN}
\bibfield{author}{\bibinfo{person}{Fan Liu}, \bibinfo{person}{Zhiyong Cheng},
  \bibinfo{person}{Lei Zhu}, \bibinfo{person}{Zan Gao}, {and}
  \bibinfo{person}{Liqiang Nie}.} \bibinfo{year}{2021}\natexlab{}.
\newblock \showarticletitle{Interest-Aware Message-Passing GCN for
  Recommendation}. In \bibinfo{booktitle}{\emph{Proceedings of the Web
  Conference 2021}}. \bibinfo{publisher}{ACM}, \bibinfo{pages}{1296–1305}.
\newblock


\bibitem[Liu et~al\mbox{.}(2024b)]%
        {ClusterGCF2024TOIS}
\bibfield{author}{\bibinfo{person}{Fan Liu}, \bibinfo{person}{Shuai Zhao},
  \bibinfo{person}{Zhiyong Cheng}, \bibinfo{person}{Liqiang Nie}, {and}
  \bibinfo{person}{Mohan Kankanhalli}.} \bibinfo{year}{2024}\natexlab{b}.
\newblock \showarticletitle{Cluster-Based Graph Collaborative Filtering}.
\newblock \bibinfo{journal}{\emph{ACM Transactions on Information Systems}}
  \bibinfo{volume}{42}, \bibinfo{number}{6} (\bibinfo{year}{2024}),
  \bibinfo{pages}{1--24}.
\newblock


\bibitem[Liu et~al\mbox{.}(2023b)]%
        {Han@MetaMMRS}
\bibfield{author}{\bibinfo{person}{Han Liu}, \bibinfo{person}{Yinwei Wei},
  \bibinfo{person}{Fan Liu}, \bibinfo{person}{Wenjie Wang},
  \bibinfo{person}{Liqiang Nie}, {and} \bibinfo{person}{Tat-Seng Chua}.}
  \bibinfo{year}{2023}\natexlab{b}.
\newblock \showarticletitle{Dynamic Multimodal Fusion via Meta-Learning Towards
  Micro-Video Recommendation}.
\newblock \bibinfo{journal}{\emph{ACM Transactions on Information Systems}}
  \bibinfo{volume}{42}, \bibinfo{number}{2} (\bibinfo{year}{2023}),
  \bibinfo{pages}{1046--8188}.
\newblock


\bibitem[Liu et~al\mbox{.}(2024a)]%
        {Liu2023Newsgeneration}
\bibfield{author}{\bibinfo{person}{Qijiong Liu}, \bibinfo{person}{Nuo Chen},
  \bibinfo{person}{Tetsuya Sakai}, {and} \bibinfo{person}{Xiao-Ming Wu}.}
  \bibinfo{year}{2024}\natexlab{a}.
\newblock \showarticletitle{ONCE: Boosting Content-based Recommendation with
  Both Open- and Closed-source Large Language Models}. In
  \bibinfo{booktitle}{\emph{Proceedings of the 17th ACM International
  Conference on Web Search and Data Mining}}. \bibinfo{publisher}{ACM},
  \bibinfo{pages}{452–461}.
\newblock


\bibitem[McAuley and Leskovec(2013)]%
        {mcauley2013hidden}
\bibfield{author}{\bibinfo{person}{Julian McAuley} {and} \bibinfo{person}{Jure
  Leskovec}.} \bibinfo{year}{2013}\natexlab{}.
\newblock \showarticletitle{Hidden factors and hidden topics: understanding
  rating dimensions with review text}. In \bibinfo{booktitle}{\emph{Proceedings
  of the 7th ACM Conference on Recommender Systems}}. \bibinfo{publisher}{ACM},
  \bibinfo{pages}{165--172}.
\newblock


\bibitem[Ren et~al\mbox{.}(2024)]%
        {ren2024representation}
\bibfield{author}{\bibinfo{person}{Xubin Ren}, \bibinfo{person}{Wei Wei},
  \bibinfo{person}{Lianghao Xia}, \bibinfo{person}{Lixin Su},
  \bibinfo{person}{Suqi Cheng}, \bibinfo{person}{Junfeng Wang},
  \bibinfo{person}{Dawei Yin}, {and} \bibinfo{person}{Chao Huang}.}
  \bibinfo{year}{2024}\natexlab{}.
\newblock \showarticletitle{Representation learning with large language models
  for recommendation}. In \bibinfo{booktitle}{\emph{Proceedings of the ACM on
  Web Conference 2024}}. \bibinfo{pages}{3464--3475}.
\newblock


\bibitem[Ren et~al\mbox{.}(2023)]%
        {Ren2023DCCF}
\bibfield{author}{\bibinfo{person}{Xubin Ren}, \bibinfo{person}{Lianghao Xia},
  \bibinfo{person}{Jiashu Zhao}, \bibinfo{person}{Dawei Yin}, {and}
  \bibinfo{person}{Chao Huang}.} \bibinfo{year}{2023}\natexlab{}.
\newblock \showarticletitle{Disentangled Contrastive Collaborative Filtering}.
  In \bibinfo{booktitle}{\emph{Proceedings of the 46th International ACM SIGIR
  Conference on Research and Development in Information Retrieval}}.
  \bibinfo{publisher}{ACM}, \bibinfo{pages}{1137–1146}.
\newblock


\bibitem[Rendle et~al\mbox{.}(2009)]%
        {rendle2009bpr}
\bibfield{author}{\bibinfo{person}{Steffen Rendle}, \bibinfo{person}{Christoph
  Freudenthaler}, \bibinfo{person}{Zeno Gantner}, {and} \bibinfo{person}{Lars
  Schmidt-Thieme}.} \bibinfo{year}{2009}\natexlab{}.
\newblock \showarticletitle{{BPR}: Bayesian personalized ranking from implicit
  feedback}. In \bibinfo{booktitle}{\emph{Proceedings of the Twenty-Fifth
  Conference on Uncertainty in Artificial Intelligence}}.
  \bibinfo{publisher}{AUAI Press}, \bibinfo{pages}{452--461}.
\newblock


\bibitem[Shuai et~al\mbox{.}(2022)]%
        {Shuai2022SIGIR}
\bibfield{author}{\bibinfo{person}{Jie Shuai}, \bibinfo{person}{Kun Zhang},
  \bibinfo{person}{Le Wu}, \bibinfo{person}{Peijie Sun},
  \bibinfo{person}{Richang Hong}, \bibinfo{person}{Meng Wang}, {and}
  \bibinfo{person}{Yong Li}.} \bibinfo{year}{2022}\natexlab{}.
\newblock \showarticletitle{A Review-aware Graph Contrastive Learning Framework
  for Recommendation}. In \bibinfo{booktitle}{\emph{Proceedings of the 45th
  International ACM SIGIR Conference on Research and Development in Information
  Retrieval}}. \bibinfo{publisher}{ACM}, \bibinfo{pages}{1283–1293}.
\newblock


\bibitem[Sileo et~al\mbox{.}(2022)]%
        {Sileo2022LLM}
\bibfield{author}{\bibinfo{person}{Damien Sileo}, \bibinfo{person}{Wout
  Vossen}, {and} \bibinfo{person}{Robbe Raymaekers}.}
  \bibinfo{year}{2022}\natexlab{}.
\newblock \showarticletitle{Zero-Shot Recommendation As Language Modeling}. In
  \bibinfo{booktitle}{\emph{Advances in Information Retrieval: 44th European
  Conference on IR Research, ECIR 2022, Stavanger, Norway, April 10–14, 2022,
  Proceedings, Part II}}. \bibinfo{publisher}{Springer-Verlag},
  \bibinfo{pages}{223–230}.
\newblock


\bibitem[Tan et~al\mbox{.}(2016)]%
        {tan2016rating}
\bibfield{author}{\bibinfo{person}{Yunzhi Tan}, \bibinfo{person}{Min Zhang},
  \bibinfo{person}{Yiqun Liu}, {and} \bibinfo{person}{Shaoping Ma}.}
  \bibinfo{year}{2016}\natexlab{}.
\newblock \showarticletitle{Rating-boosted latent topics: Understanding users
  and items with ratings and reviews}. In \bibinfo{booktitle}{\emph{Proceedings
  of the Twenty-Fifth International Joint Conference on Artificial
  Intelligence}}. \bibinfo{publisher}{AAAI Press}, \bibinfo{pages}{2640--2646}.
\newblock


\bibitem[van~den Berg et~al\mbox{.}(2018)]%
        {berg2019gcmc}
\bibfield{author}{\bibinfo{person}{Rianne van~den Berg},
  \bibinfo{person}{Thomas~N. Kipf}, {and} \bibinfo{person}{Max Welling}.}
  \bibinfo{year}{2018}\natexlab{}.
\newblock \showarticletitle{Graph Convolutional Matrix Completion}. In
  \bibinfo{booktitle}{\emph{ACM SIGKDD: Deep Learning Day}}.
  \bibinfo{publisher}{ACM}.
\newblock


\bibitem[Wang et~al\mbox{.}(2019a)]%
        {wang2019kdd}
\bibfield{author}{\bibinfo{person}{Xiang Wang}, \bibinfo{person}{Xiangnan He},
  \bibinfo{person}{Yixin Cao}, \bibinfo{person}{Meng Liu}, {and}
  \bibinfo{person}{Tat-Seng Chua}.} \bibinfo{year}{2019}\natexlab{a}.
\newblock \showarticletitle{{KGAT:} Knowledge Graph Attention Network for
  Recommendation}. In \bibinfo{booktitle}{\emph{Proceedings of the 25th ACM
  SIGKDD International Conference on Knowledge Discovery \& Data Mining}}.
  \bibinfo{publisher}{ACM}, \bibinfo{pages}{950--958}.
\newblock


\bibitem[Wang et~al\mbox{.}(2019b)]%
        {wang2019ngcf}
\bibfield{author}{\bibinfo{person}{Xiang Wang}, \bibinfo{person}{Xiangnan He},
  \bibinfo{person}{Meng Wang}, \bibinfo{person}{Fuli Feng}, {and}
  \bibinfo{person}{Tat-Seng Chua}.} \bibinfo{year}{2019}\natexlab{b}.
\newblock \showarticletitle{Neural Graph Collaborative Filtering}. In
  \bibinfo{booktitle}{\emph{Proceedings of the 42nd International ACM SIGIR
  Conference on Research and Development in Information Retrieval}}.
  \bibinfo{publisher}{ACM}, \bibinfo{pages}{165--174}.
\newblock


\bibitem[Wang et~al\mbox{.}(2020a)]%
        {wang2020DGCF}
\bibfield{author}{\bibinfo{person}{Xiang Wang}, \bibinfo{person}{Hongye Jin},
  \bibinfo{person}{An Zhang}, \bibinfo{person}{Xiangnan He},
  \bibinfo{person}{Tong Xu}, {and} \bibinfo{person}{Tat-Seng Chua}.}
  \bibinfo{year}{2020}\natexlab{a}.
\newblock \showarticletitle{Disentangled Graph Collaborative Filtering}. In
  \bibinfo{booktitle}{\emph{Proceedings of the 43rd International ACM SIGIR
  Conference on Research and Development in Information Retrieval}}.
  \bibinfo{publisher}{ACM}, \bibinfo{pages}{1001–1010}.
\newblock


\bibitem[Wang et~al\mbox{.}(2020b)]%
        {wang2020DisenHAN}
\bibfield{author}{\bibinfo{person}{Yifan Wang}, \bibinfo{person}{Suyao Tang},
  \bibinfo{person}{Yuntong Lei}, \bibinfo{person}{Weiping Song},
  \bibinfo{person}{Sheng Wang}, {and} \bibinfo{person}{Ming Zhang}.}
  \bibinfo{year}{2020}\natexlab{b}.
\newblock \showarticletitle{DisenHAN: Disentangled Heterogeneous Graph
  Attention Network for Recommendation}. In
  \bibinfo{booktitle}{\emph{Proceedings of the 2020 ACM on Conference on
  Information and Knowledge Management}}. \bibinfo{publisher}{ACM},
  \bibinfo{pages}{1605–1614}.
\newblock


\bibitem[Wei et~al\mbox{.}(2019)]%
        {wei2019mm}
\bibfield{author}{\bibinfo{person}{Yinwei Wei}, \bibinfo{person}{Zhiyong
  Cheng}, \bibinfo{person}{Xuzheng Yu}, \bibinfo{person}{Zhou Zhao},
  \bibinfo{person}{Lei Zhu}, {and} \bibinfo{person}{Liqiang Nie}.}
  \bibinfo{year}{2019}\natexlab{}.
\newblock \showarticletitle{Personalized Hashtag Recommendation for
  Micro-videos}. In \bibinfo{booktitle}{\emph{Proceedings of the 27th ACM
  International Conference on Multimedia}}. \bibinfo{publisher}{ACM},
  \bibinfo{pages}{1446--1454}.
\newblock


\bibitem[Wei et~al\mbox{.}(2023)]%
        {Wei2023LightGT}
\bibfield{author}{\bibinfo{person}{Yinwei Wei}, \bibinfo{person}{Wenqi Liu},
  \bibinfo{person}{Fan Liu}, \bibinfo{person}{Xiang Wang},
  \bibinfo{person}{Liqiang Nie}, {and} \bibinfo{person}{Tat-Seng Chua}.}
  \bibinfo{year}{2023}\natexlab{}.
\newblock \showarticletitle{LightGT: A Light Graph Transformer for Multimedia
  Recommendation}. In \bibinfo{booktitle}{\emph{Proceedings of the 46th
  International ACM SIGIR Conference on Research and Development in Information
  Retrieval}}. \bibinfo{publisher}{ACM}, \bibinfo{pages}{1508–1517}.
\newblock


\bibitem[Wei et~al\mbox{.}(2020)]%
        {Wei2019GRCN}
\bibfield{author}{\bibinfo{person}{Yinwei Wei}, \bibinfo{person}{Xiang Wang},
  \bibinfo{person}{Liqiang Nie}, \bibinfo{person}{Xiangnan He}, {and}
  \bibinfo{person}{Tat-Seng Chua}.} \bibinfo{year}{2020}\natexlab{}.
\newblock \showarticletitle{Graph-Refined Convolutional Network for Multimedia
  Recommendation with Implicit Feedback}. In
  \bibinfo{booktitle}{\emph{Proceedings of the 28th ACM International
  Conference on Multimedia}}. \bibinfo{publisher}{ACM},
  \bibinfo{pages}{3541--3549}.
\newblock


\bibitem[Wu et~al\mbox{.}(2019)]%
        {pmlr-v97-wu19e}
\bibfield{author}{\bibinfo{person}{Felix Wu}, \bibinfo{person}{Amauri Souza},
  \bibinfo{person}{Tianyi Zhang}, \bibinfo{person}{Christopher Fifty},
  \bibinfo{person}{Tao Yu}, {and} \bibinfo{person}{Kilian Weinberger}.}
  \bibinfo{year}{2019}\natexlab{}.
\newblock \showarticletitle{Simplifying Graph Convolutional Networks}. In
  \bibinfo{booktitle}{\emph{Proceedings of the 36th International Conference on
  Machine Learning}}. \bibinfo{publisher}{PMLR}, \bibinfo{pages}{6861--6871}.
\newblock


\bibitem[Wu et~al\mbox{.}(2016)]%
        {wu2016cdae}
\bibfield{author}{\bibinfo{person}{Yao Wu}, \bibinfo{person}{Christopher
  DuBois}, \bibinfo{person}{Alice~X. Zheng}, {and} \bibinfo{person}{Martin
  Ester}.} \bibinfo{year}{2016}\natexlab{}.
\newblock \showarticletitle{Collaborative Denoising Auto-Encoders for Top-N
  Recommender Systems}. In \bibinfo{booktitle}{\emph{Proceedings of the Tenth
  ACM International Conference on Web Search and Data Mining}}.
  \bibinfo{publisher}{ACM}, \bibinfo{pages}{153--162}.
\newblock


\bibitem[Xue et~al\mbox{.}(2017)]%
        {xue2017deep}
\bibfield{author}{\bibinfo{person}{HongJian Xue}, \bibinfo{person}{XinYu Dai},
  \bibinfo{person}{Jianbing Zhang}, \bibinfo{person}{Shujian Huang}, {and}
  \bibinfo{person}{Jiajun Chen}.} \bibinfo{year}{2017}\natexlab{}.
\newblock \showarticletitle{Deep matrix factorization models for recommender
  systems}. In \bibinfo{booktitle}{\emph{Proceedings of the 26th International
  Joint Conference on Artificial Intelligence}}. \bibinfo{publisher}{AAAI
  Press}, \bibinfo{pages}{3203--3209}.
\newblock


\bibitem[Yang et~al\mbox{.}(2018)]%
        {hoprec}
\bibfield{author}{\bibinfo{person}{Jheng{-}Hong Yang},
  \bibinfo{person}{Chih{-}Ming Chen}, \bibinfo{person}{Chuan{-}Ju Wang}, {and}
  \bibinfo{person}{Ming{-}Feng Tsai}.} \bibinfo{year}{2018}\natexlab{}.
\newblock \showarticletitle{{HOP}-rec: high-order proximity for implicit
  recommendation}. In \bibinfo{booktitle}{\emph{Proceedings of the 12th ACM
  Conference on Recommender Systems}}. \bibinfo{publisher}{ACM},
  \bibinfo{pages}{140--144}.
\newblock


\bibitem[Ying et~al\mbox{.}(2018)]%
        {ying2018pinsage}
\bibfield{author}{\bibinfo{person}{Rex Ying}, \bibinfo{person}{Ruining He},
  \bibinfo{person}{Kaifeng Chen}, \bibinfo{person}{Pong Eksombatchai},
  \bibinfo{person}{William~L. Hamilton}, {and} \bibinfo{person}{Jure
  Leskovec}.} \bibinfo{year}{2018}\natexlab{}.
\newblock \showarticletitle{Graph Convolutional Neural Networks for Web-Scale
  Recommender Systems}. In \bibinfo{booktitle}{\emph{Proceedings of the 24th
  ACM SIGKDD International Conference on Knowledge Discovery \& Data Mining}}.
  \bibinfo{publisher}{ACM}, \bibinfo{pages}{974--983}.
\newblock


\bibitem[Yu et~al\mbox{.}(2018)]%
        {yu2018walkranker}
\bibfield{author}{\bibinfo{person}{Lu Yu}, \bibinfo{person}{Chuxu Zhang},
  \bibinfo{person}{Shichao Pei}, \bibinfo{person}{Guolei Sun}, {and}
  \bibinfo{person}{Xiangliang Zhang}.} \bibinfo{year}{2018}\natexlab{}.
\newblock \showarticletitle{WalkRanker: {A} Unified Pairwise Ranking Model With
  Multiple Relations for Item Recommendation}. In
  \bibinfo{booktitle}{\emph{Proceedings of the Thirty-Second AAAI Conference on
  Artificial Intelligence and Thirtieth Innovative Applications of Artificial
  Intelligence Conference and Eighth AAAI Symposium on Educational Advances in
  Artificial Intelligence}}. \bibinfo{publisher}{{AAAI} Press},
  \bibinfo{pages}{2596--2603}.
\newblock


\bibitem[Yue et~al\mbox{.}(2023)]%
        {yue2023llamarec}
\bibfield{author}{\bibinfo{person}{Zhenrui Yue}, \bibinfo{person}{Sara Rabhi},
  \bibinfo{person}{Gabriel de Souza~Pereira Moreira}, \bibinfo{person}{Dong
  Wang}, {and} \bibinfo{person}{Even Oldridge}.}
  \bibinfo{year}{2023}\natexlab{}.
\newblock \bibinfo{title}{LlamaRec: Two-Stage Recommendation using Large
  Language Models for Ranking}.
\newblock
\newblock
\showeprint{2311.02089}


\bibitem[Zhang et~al\mbox{.}(2016)]%
        {zhang2016collaborative}
\bibfield{author}{\bibinfo{person}{Fuzheng Zhang},
  \bibinfo{person}{Nicholas~Jing Yuan}, \bibinfo{person}{Defu Lian},
  \bibinfo{person}{Xing Xie}, {and} \bibinfo{person}{Wei-Ying Ma}.}
  \bibinfo{year}{2016}\natexlab{}.
\newblock \showarticletitle{Collaborative knowledge base embedding for
  recommender systems}. In \bibinfo{booktitle}{\emph{Proceedings of the 22nd
  ACM SIGKDD International Conference on Knowledge Discovery and Data Mining}}.
  ACM, \bibinfo{pages}{353--362}.
\newblock


\bibitem[Zhang et~al\mbox{.}(2023a)]%
        {Zhang2023chatgpt}
\bibfield{author}{\bibinfo{person}{Jizhi Zhang}, \bibinfo{person}{Keqin Bao},
  \bibinfo{person}{Yang Zhang}, \bibinfo{person}{Wenjie Wang},
  \bibinfo{person}{Fuli Feng}, {and} \bibinfo{person}{Xiangnan He}.}
  \bibinfo{year}{2023}\natexlab{a}.
\newblock \showarticletitle{Is ChatGPT Fair for Recommendation? Evaluating
  Fairness in Large Language Model Recommendation}. In
  \bibinfo{booktitle}{\emph{Proceedings of the 17th ACM Conference on
  Recommender Systems}}. \bibinfo{publisher}{ACM}, \bibinfo{pages}{993–999}.
\newblock


\bibitem[Zhang et~al\mbox{.}(2023b)]%
        {Zhang2023LLM}
\bibfield{author}{\bibinfo{person}{Junjie Zhang}, \bibinfo{person}{Ruobing
  Xie}, \bibinfo{person}{Yupeng Hou}, \bibinfo{person}{Wayne~Xin Zhao},
  \bibinfo{person}{Leyu Lin}, {and} \bibinfo{person}{Ji{-}Rong Wen}.}
  \bibinfo{year}{2023}\natexlab{b}.
\newblock \bibinfo{title}{Recommendation as Instruction Following: {A} Large
  Language Model Empowered Recommendation Approach}.
\newblock
\newblock
\showeprint{2305.07001}


\bibitem[Zhang et~al\mbox{.}(2021)]%
        {LATTICE2021TMM}
\bibfield{author}{\bibinfo{person}{Jinghao Zhang}, \bibinfo{person}{Yanqiao
  Zhu}, \bibinfo{person}{Qiang Liu}, \bibinfo{person}{Shu Wu},
  \bibinfo{person}{Shuhui Wang}, {and} \bibinfo{person}{Liang Wang}.}
  \bibinfo{year}{2021}\natexlab{}.
\newblock \showarticletitle{Mining Latent Structures for Multimedia
  Recommendation}. In \bibinfo{booktitle}{\emph{Proceedings of the 29th ACM
  International Conference on Multimedia}}. \bibinfo{publisher}{ACM},
  \bibinfo{pages}{3872–3880}.
\newblock


\bibitem[Zhang et~al\mbox{.}(2017)]%
        {zhang2017joint}
\bibfield{author}{\bibinfo{person}{Yongfeng Zhang}, \bibinfo{person}{Qingyao
  Ai}, \bibinfo{person}{Xu Chen}, {and} \bibinfo{person}{W~Bruce Croft}.}
  \bibinfo{year}{2017}\natexlab{}.
\newblock \showarticletitle{Joint representation learning for top-n
  recommendation with heterogeneous information sources}. In
  \bibinfo{booktitle}{\emph{Proceedings of the 2017 ACM on Conference on
  Information and Knowledge Management}}. \bibinfo{publisher}{ACM},
  \bibinfo{pages}{1449--1458}.
\newblock


\bibitem[Zheng et~al\mbox{.}(2017)]%
        {DeepCoNN2017WSDM}
\bibfield{author}{\bibinfo{person}{Lei Zheng}, \bibinfo{person}{Vahid Noroozi},
  {and} \bibinfo{person}{Philip~S. Yu}.} \bibinfo{year}{2017}\natexlab{}.
\newblock \showarticletitle{Joint Deep Modeling of Users and Items Using
  Reviews for Recommendation}. In \bibinfo{booktitle}{\emph{Proceedings of the
  Tenth ACM International Conference on Web Search and Data Mining}}.
  \bibinfo{publisher}{ACM}, \bibinfo{pages}{425–434}.
\newblock


\bibitem[Zhou et~al\mbox{.}(2023)]%
        {BM32020Arxiv}
\bibfield{author}{\bibinfo{person}{Xin Zhou}, \bibinfo{person}{Hongyu Zhou},
  \bibinfo{person}{Yong Liu}, \bibinfo{person}{Zhiwei Zeng},
  \bibinfo{person}{Chunyan Miao}, \bibinfo{person}{Pengwei Wang},
  \bibinfo{person}{Yuan You}, {and} \bibinfo{person}{Feijun Jiang}.}
  \bibinfo{year}{2023}\natexlab{}.
\newblock \showarticletitle{Bootstrap Latent Representations for Multi-modal
  Recommendation}. In \bibinfo{booktitle}{\emph{Proceedings of the ACM Web
  Conference 2023}}. \bibinfo{publisher}{ACM}, \bibinfo{pages}{845–854}.
\newblock


\end{thebibliography}

\end{document}